\documentclass[runningheads]{llncs}
\usepackage[utf8]{inputenc}
\usepackage[T1]{fontenc}

\usepackage[british]{babel}

\usepackage[numbers]{natbib}

\usepackage{amsmath}
\usepackage{amssymb}

\usepackage{marvosym} % Letter symbol

\usepackage{mathpartir}

\usepackage[colorlinks=true,linkcolor=black,citecolor=black,filecolor=black,urlcolor=black,pdfpagelabels]{hyperref}

\usepackage{isabelle,isabellesym}
\isabellestyle{it}

\isadroptag{theory}
\isafoldtag{proof}

\usepackage[capitalise]{cleveref}

\usepackage{xcolor}

\usepackage{tikz}
\usetikzlibrary{shapes,positioning,patterns,patterns.meta,trees}
\usepackage{pgfplots}
\pgfplotsset{compat=1.18}
\pgfdeclarelayer{background}
\pgfdeclarelayer{foreground}
\pgfsetlayers{background,main,foreground}

\usepackage{calc}

\usepackage[inline]{enumitem}

\usepackage{array}
\usepackage{multirow}

\usepackage{booktabs}

\usepackage{subcaption}

\usepackage{todonotes}

\makeatletter
\renewcommand\subsubsection{\@startsection{subsubsection}{3}{\z@}%
                       {-8\p@ \@plus -4\p@ \@minus -4\p@}% Formerly -18\p@ \@plus -4\p@ \@minus -4\p@
                       {-0.5em \@plus -0.22em \@minus -0.1em}%
                       {\normalfont\normalsize\bfseries\boldmath}}
\makeatother

\usepackage[acronym,section=section]{glossaries}
\makenoidxglossaries
\newacronym{adt}{ADT}{abstract data type}
\newacronym{uf}{UF}{Union-Find}
\newacronym{ufe}{UFE}{Union-Find-Explain}
\newacronym{lca}{LCA}{lowest common ancestor}
\newacronym{afp}{AFP}{Archive of Formal Proofs}
\newacronym{smt}{SMT}{Satisfiability Modulo Theories}
\newacronym{sml}{SML}{Standard ML}
\newacronym{tcb}{TCB}{trusted computing base}

\newcommand{\opunion}{\textsc{union}}
\newcommand{\opfind}{\textsc{find}}
\newcommand{\opexplain}{\textsc{explain}}

\title{Simplified and Verified: A Second Look at a Proof-Producing Union-Find Algorithm}
\titlerunning{Verifying a Proof-Producing Union-Find Algorithm}
\author{Lukas Stevens\inst{1,2}\orcidID{0000-0003-0222-6858} \and\\ Rebecca Ghidini\inst{1,3,4,5}\orcidID{0009-0009-8117-2659}}
\institute{
  Technical University of Munich, Department of Computer Science,\\
  Boltzmannstr. 3, Garching, Germany
  \and
  \email{lukas.stevens@in.tum.de} (\Letter)
  \and
  École normale supérieure, Département d’informatique de l’ENS,\\
  CNRS, PSL University, 75005 Paris, France
  \and
  Inria
  \and
  \email{rebecca.ghidini@info.ens.psl.eu}
}

\begin{document}

\begin{isabellebody}%
\setisabellecontext{LaTeXsugar}%
\isadelimtheory
\isanewline
\endisadelimtheory
\isatagtheory
\endisatagtheory
{\isafoldtheory}%
\isadelimtheory
\endisadelimtheory
\isadelimML
\endisadelimML
\isatagML
\endisatagML
{\isafoldML}%
\isadelimML
\endisadelimML
\isadelimML
\endisadelimML
\isatagML
\endisatagML
{\isafoldML}%
\isadelimML
\endisadelimML
\isadelimML
\endisadelimML
\isatagML
\endisatagML
{\isafoldML}%
\isadelimML
\endisadelimML
\isadelimtheory
\endisadelimtheory
\isatagtheory
\endisatagtheory
{\isafoldtheory}%
\isadelimtheory
\endisadelimtheory
\end{isabellebody}%

\begin{isabellebody}%
\setisabellecontext{Document}%
\isadelimtheory
\endisadelimtheory
\isatagtheory
\isakeywordONE{theory}\isamarkupfalse%
\ Document\isanewline
\ \ \isakeywordTWO{imports}\ Main\ {\isachardoublequoteopen}Union{\isacharunderscore}{\kern0pt}Find{\isacharunderscore}{\kern0pt}Explain{\isachardot}{\kern0pt}Explain{\isacharunderscore}{\kern0pt}Imp{\isachardoublequoteclose}\ LaTeXsugar\ \isanewline
\isakeywordTWO{begin}%
\endisatagtheory
{\isafoldtheory}%
\isadelimtheory
\endisadelimtheory
\renewcommand{\isakeyword}[1]{\ensuremath{\mathsf{#1}}}
\renewcommand{\isacommand}[1]{\isakeywordONE{#1}}
\renewcommand{\isaconst}[1]{\ensuremath{\mathsf{#1}}}
\newlength{\funheadersep}
\setlength{\funheadersep}{2pt}
\setenumerate[0]{label=(\arabic*)}
\newcommand{\Cpp}[0]{C\texttt{++}}
\newcommand{\Gpp}[0]{g\texttt{++}}
\begin{isamarkuptext}%
\maketitle%
\end{isamarkuptext}\isamarkuptrue%
\begin{isamarkuptext}%
\begin{abstract}
Using Isabelle/HOL, we verify a union-find data structure with an explain operation due to \citeauthor{congcl_proofs}.
We devise a simpler, more naive version of the explain operation whose soundness and completeness is easy to verify.
Then, we prove the original formulation of the explain operation to be equal to our version.
Finally, we refine this data structure to Imperative HOL, enabling us to export efficient imperative code.
The formalisation provides a stepping stone towards the verification of proof-producing congruence closure algorithms which are a core ingredient of \acrfull{smt} solvers.

\keywords{Equivalence closure \and Interactive theorem proving \and Satisfiability modulo theories \and Proof-producing decision procedure}
\end{abstract}%
\end{isamarkuptext}\isamarkuptrue%
\begin{isamarkuptext}%
%\todo[inline]{Move question marks to ROOT}
%\todo[inline]{Rename au\_ds to au and uf\_ds to uf?}
%\todo[inline]{Use cref when sensible (e.g.\ when referring to current section)}
%\todo[inline]{Think about \isa{{\isacharparenleft}{\kern0pt}{\isacharcolon}{\kern0pt}{\isacharequal}{\kern0pt}{\isacharparenright}{\kern0pt}} vs \isa{{\isacharparenleft}{\kern0pt}{\isasymequiv}{\isacharparenright}{\kern0pt}}}%
\end{isamarkuptext}\isamarkuptrue%
\isadelimdocument
\endisadelimdocument
\isatagdocument
\isamarkupsection{Introduction%
}
\isamarkuptrue%
\endisatagdocument
{\isafolddocument}%
\isadelimdocument
\endisadelimdocument
\begin{isamarkuptext}%
The \acrfull{uf} data structure maintains the equivalence closure of a relation,
which is given as a sequence of pairs or, in terms of the \acrshort{uf} data structure, \opunion{} operations.
It is fundamental to efficiently implement well-known graph algorithms such as \citeauthor{mst}'s~\cite{mst} minimum spanning-tree algorithm. 
There, it tracks which vertices belong to the same connected component and are, in that sense, equivalent.
Beyond graph algorithms, its applicability extends to the domain of theorem proving as it routinely forms the basis of congruence closure algorithms, which are widely used by \acrfull{smt} solvers.
To increase their trustworthiness, current \acrshort{smt} solvers such as
cvc5~\cite{cvc5}, E~\cite{eprover}, Vampire~\cite{vampire}, veriT~\cite{verit}, and Z3~\cite{z3_proofs}
can output detailed proofs when they determine an input formula to be unsatisfiable.
To produce these proofs, it is crucial to have congruence closure algorithms that efficiently explain why they consider two terms to be equal.
The first such algorithm was presented by \citeauthor{congcl_proofs}~\cite{congcl_proofs,congcl_fast_extensions}, 
who extended the \acrshort{uf} data structure with an \opexplain{} operation to obtain a \acrfull{ufe} data structure.
Verifying this data structure in Isabelle/HOL is the focus of our paper.

Why, then, should we verify a data structure that already produces proofs?
Our answer is three-fold.
\begin{enumerate*}[label=(\arabic*)]
\item While the data structure's proofs guarantee soundness, we also prove its completeness.
\item Executing code exported from Isabelle depends on a substantial \acrfull{tcb}, including the target language's compiler and runtime.
\item Whereas \acrshort{smt} solvers typically operate with a large \acrshort{tcb},
adding to the \acrshort{tcb} of interactive theorem provers is usually discouraged.
Proof-producing algorithms --- such as those introduced in earlier work by the first author~\cite{orders} --- enable integration without increasing the \acrshort{tcb}.
\end{enumerate*}%
\end{isamarkuptext}\isamarkuptrue%
\isadelimdocument
\endisadelimdocument
\isatagdocument
\isamarkupsubsection{Contributions%
}
\isamarkuptrue%
\endisatagdocument
{\isafolddocument}%
\isadelimdocument
\endisadelimdocument
\begin{isamarkuptext}%
We present, to our knowledge, the first formalisation of the \acrshort{ufe} data structure as introduced by \citeauthor{congcl_proofs}.
In their work, they present two variants of this data structure.
Here, we only formalise the first variant, leaving the other for future work.
We devise a simpler, more naive version of the \opexplain{} operation, for which soundness and completeness is easier to prove. 
Then, we prove the original version of the \opexplain{} operation to be extensionally equal to the simple one.
Based on an existing formalisation of the \acrshort{uf} data structure by \citeauthor{uf_isabelle}~\cite{uf_isabelle},
we develop a more abstract formalisation of the data structure, hiding implementation details. 
Finally, we refine the \acrshort{ufe} data structure to Imperative HOL~\cite{imperative_hol} using \citeauthor{uf_isabelle}'s~\cite{uf_isabelle} separation logic framework,
enabling generation of efficient imperative code.

The formalisation is available online.\footnote{\url{https://doi.org/10.5281/zenodo.15557955}}
Since everything is verified, we omit proofs and focus on outlining the structure of the formalisation.%
\end{isamarkuptext}\isamarkuptrue%
\isadelimdocument
\endisadelimdocument
\isatagdocument
\isamarkupsubsection{Related Work%
}
\isamarkuptrue%
\endisatagdocument
{\isafolddocument}%
\isadelimdocument
\endisadelimdocument
\begin{isamarkuptext}%
Efficient implementations of the \acrshort{uf} data structure have been known for a long time.
In particular, \citeauthor{uf_by_size}~\cite{uf_by_size} represent the data structure as a forest of rooted trees
and propose the union-by-size heuristic,
which gives $\mathcal{O}(\log n)$ running time for \opunion{} and \opfind{} for a data struture over $n$ elements.
Another heuristic, called path compression, was presented by \citeauthor{uf_compress}~\cite{uf_compress}.
Analysing the complexity of the data structure when combining both heuristics turned out to be challenging,
but \citeauthor{uf_ub}~\cite{uf_ub} and \citeauthor{uf_ub_improved}~\cite{uf_ub_improved}
eventually proved an amortised running time of $\mathcal{O}(n + m\, \alpha(m + n, n))$
for a sequence of at most $n - 1$ \opunion{} and $m$ \opfind{} operations where $\alpha$ is the inverse Ackermann function. 
This means that any one operation runs in almost constant time, amortised.

While the paper on the \acrshort{ufe} data structure~\cite{congcl_proofs} is widely cited,
there is limited follow-up literature on this data structure.
It does, however, form the basis of proof-producing congruence closure algorithms, which are crucial in the field of \acrshort{smt} solving.
There, they remain an active area of research;
for example, when we are interested in efficiently finding small proofs~\cite{congcl_small_proofs}.

The literature of verified algorithms and data structures is vast so we refer to a survey~\cite{algorithms_survey} for an overview.
Focusing on the \acrshort{uf} data structure, there is a formalisation in Coq~\cite{uf_coq}, where
the amortised complexity is analysed by \citeauthor{uf_coq_time}~\cite{uf_coq_time} in a separation logic with time credits.
Similarly, in Isabelle, there is a formalisation of the data structure~\cite{uf_isabelle}
that was later extended with a complexity analysis by \citeauthor{uf_isabelle_time}~\cite{uf_isabelle_time}.
More recently, there is formalisation by \citeauthor{uf_isabelle_algebraic}~\cite{uf_isabelle_algebraic} taking a relation-algebraic view.%
\end{isamarkuptext}\isamarkuptrue%
\isadelimdocument
\endisadelimdocument
\isatagdocument
\isamarkupsubsection{Notation%
}
\isamarkuptrue%
\endisatagdocument
{\isafolddocument}%
\isadelimdocument
\endisadelimdocument
\begin{isamarkuptext}%
Isabelle/HOL~\cite{isabelle} conforms to everyday mathematical notation for the most part.
We establish notation and in particular some essential data types together with their primitive operations that are specific to Isabelle/HOL.

We write \isa{\isafree{t}{\isacharcolon}{\kern0pt}{\isacharcolon}{\kern0pt}\isatfree{{\isacharprime}{\kern0pt}a}} to specify that the term \isa{\isafree{t}} has the type \isa{\isatfree{{\isacharprime}{\kern0pt}a}} and \isa{\isatfree{{\isacharprime}{\kern0pt}a}\ {\isasymRightarrow}\ \isatfree{{\isacharprime}{\kern0pt}b}} for the space of total functions from type \isa{\isatfree{{\isacharprime}{\kern0pt}a}} to type \isa{\isatfree{{\isacharprime}{\kern0pt}b}}.

Sets with elements of type \isa{\isatfree{{\isacharprime}{\kern0pt}a}} have the type \isa{\isatfree{{\isacharprime}{\kern0pt}a}\ \isatconst{set}}.
The cardinality of a set \isa{\isafree{A}} is denoted by \isa{{\isacharbar}{\kern0pt}\isafree{A}{\isacharbar}{\kern0pt}}.

We use \isa{\isatfree{{\isacharprime}{\kern0pt}a}\ \isatconst{list}} to describe the type of lists, which are constructed using the empty list \isa{{\isacharbrackleft}{\kern0pt}{\isacharbrackright}{\kern0pt}}
or the infix cons constructor \isa{{\isacharparenleft}{\kern0pt}{\isacharhash}{\kern0pt}{\isacharparenright}{\kern0pt}},
and are appended with the infix operator \isa{{\isacharparenleft}{\kern0pt}{\isacharat}{\kern0pt}{\isacharparenright}{\kern0pt}}.
The length of list \isa{\isafree{xs}} is denote by \isa{{\isacharbar}{\kern0pt}\isafree{xs}{\isacharbar}{\kern0pt}}.
The function \isa{\isaconst{set}} converts a list into a set.
We write \isa{\isafree{xs}\ {\isacharbang}{\kern0pt}\ \isafree{i}} to access the \isa{\isafree{i}}-th element of the list \isa{\isafree{xs}}.

To represent partial values of type \isa{\isatfree{{\isacharprime}{\kern0pt}a}}, we use the type \isa{\isatfree{{\isacharprime}{\kern0pt}a}\ \isatconst{option}} with the constructors \isa{\isaconst{None}} and \isa{\isaconst{Some}}.
We also define an order on this type by letting \isa{\isaconst{None}} be smaller than \isa{\isaconst{Some}}
and lifting the order on the underlying type \isa{\isatfree{{\isacharprime}{\kern0pt}a}},
i.e.\ we have that \isa{{\isacharparenleft}{\kern0pt}\isaconst{Some}\ \isafree{x}\ {\isasymle}\ \isaconst{Some}\ \isafree{y}{\isacharparenright}{\kern0pt}\ {\isacharequal}{\kern0pt}\ {\isacharparenleft}{\kern0pt}\isafree{x}\ {\isasymle}\ \isafree{y}{\isacharparenright}{\kern0pt}}.

Relations are denoted with the type synonym \isa{\isatfree{{\isacharprime}{\kern0pt}a}\ \isatconst{rel}}, which expands to \isa{{\isacharparenleft}{\kern0pt}\isatfree{{\isacharprime}{\kern0pt}a}\ {\isasymtimes}\ \isatfree{{\isacharprime}{\kern0pt}a}{\isacharparenright}{\kern0pt}\ \isatconst{set}}.
For a relation \isa{\isafree{r}}, \isa{\isaconst{Field}\ \isafree{r}} are those elements that occur as a component of a pair \isa{\isafree{p}\ {\isasymin}\ \isafree{r}}.
Furthermore, we use \isa{\isafree{r}{\isasyminverse}} to denote the inverse and \isa{\isafree{r}\isactrlsup {\isacharasterisk}{\kern0pt}} to denote the reflexive transitive closure of \isa{\isafree{r}}.

Throughout our formalisation we employ \emph{locales}~\cite{locales},
which are named contexts of types, constants and assumptions about them.%
\end{isamarkuptext}\isamarkuptrue%
\isadelimdocument
\endisadelimdocument
\isatagdocument
\isamarkupsection{Basic Union-Find%
}
\isamarkuptrue%
\isamarkupsubsection{Background\label{sec:uf_background}%
}
\isamarkuptrue%
\endisatagdocument
{\isafolddocument}%
\isadelimdocument
\endisadelimdocument
\begin{isamarkuptext}%
Given a set of $n$ elements $A = \{a_1, \ldots, a_n\}$, the \acrshort{uf} data structure keeps track of a partition of $A$ into disjoint sets $A_1, \ldots, A_k$, i.e.\ $A = A_1 \uplus \cdots \uplus A_k$.
Equivalently, one can view the partition as a partial equivalence relation with the equivalence classes $A_1, \ldots, A_k$.
The equivalence relation is partial because \isa{\isafree{A}{\isacharcolon}{\kern0pt}{\isacharcolon}{\kern0pt}\isatfree{{\isacharprime}{\kern0pt}a}\ \isatconst{set}} might only be a subset of the type \isa{\isatfree{{\isacharprime}{\kern0pt}a}}.
We initialise the data structure by partitioning $A$ into singleton sets of elements,
so we have that $A = \{a_1\} \uplus \cdots \uplus \{a_n\}$.
Those sets are merged by subsequent \opunion{} operations where $\opunion{}~a_i~a_j$ merges the set containing $a_i$ with the one that contains $a_j$.
Each set in the partition contains one particular element that serves as its representative.
We will denote the representative of an element \isa{\isafree{a}}
in the \acrshort{uf} data structure \isa{\isafree{uf}}
as \isa{\isaconst{rep{\isacharunderscore}{\kern0pt}of}\ \isafree{uf}\ \isafree{a}}.
Accordingly, two elements have the same representative exactly when they belong to the same set in the partition.
For any element \isa{\isafree{a\isactrlsub i}}, the \opfind{} operation returns its representative \isa{\isaconst{rep{\isacharunderscore}{\kern0pt}of}\ \isafree{uf}\ \isafree{a\isactrlsub i}}.

The data structure can be implemented as a forest of rooted trees
where each tree encodes an equivalence class.
The edges of a tree in the forest are directed towards the root,
which is the representative of the corresponding equivalence class.
To preserve this invariant, we initialise the forest with $n$ vertices but without any edges
and, for every \opunion{} of $a_i$ and $a_j$,
we add a directed edge from \isa{\isaconst{rep{\isacharunderscore}{\kern0pt}of}\ \isafree{uf}\ \isafree{a\isactrlsub i}} to \isa{\isaconst{rep{\isacharunderscore}{\kern0pt}of}\ \isafree{uf}\ \isafree{a\isactrlsub j}} to the forest.

We encode such a forest as a list \isa{\isafree{l}} of length \isa{\isafree{n}},
where at each index \isa{\isafree{i}} of \isa{\isafree{l}}, we save the index of the parent of the element \isa{\isafree{a\isactrlsub i}}, denoted by \isa{\isafree{l}\ {\isacharbang}{\kern0pt}\ \isafree{i}}.
If \isa{\isafree{a\isactrlsub i}} is a root, then the list stores \isa{\isafree{i}} itself at index \isa{\isafree{i}},
i.e. \isa{\isafree{l}\ {\isacharbang}{\kern0pt}\ \isafree{i}\ {\isacharequal}{\kern0pt}\ \isafree{i}}.%
\end{isamarkuptext}\isamarkuptrue%
\isadelimdocument
\endisadelimdocument
\isatagdocument
\isamarkupsubsection{In Isabelle/HOL\label{sec:uf_hol}%
}
\isamarkuptrue%
\endisatagdocument
{\isafolddocument}%
\isadelimdocument
\endisadelimdocument
\begin{isamarkuptext}%
The \acrshort{uf} algorithm was formalised in Isabelle/HOL by \citeauthor{uf_isabelle}~\cite{uf_isabelle}.
The code can be found in an entry~\cite{uf_isabelle_afp} of the \acrfull{afp}.\footnote{The code is in the theory file \texttt{Examples/Union\_Find.thy}.}
\citeauthor{uf_isabelle} defines a function \isa{\isaconst{rep{\isacharunderscore}{\kern0pt}of}},
which, as described above, follows the parent pointers until we arrive at the root,
where the parent pointer is self-referential.
\begin{flushleft}
{\parindent0pt\isa{\isaconst{rep{\isacharunderscore}{\kern0pt}of}\ {\isacharcolon}{\kern0pt}{\isacharcolon}{\kern0pt}\ \isatconst{nat}\ \isatconst{list}\ {\isasymRightarrow}\ \isatconst{nat}\ {\isasymRightarrow}\ \isatconst{nat}}\\[\funheadersep]\isa{\isaconst{rep{\isacharunderscore}{\kern0pt}of}\ \isafree{l}\ \isavar{i}\ {\isacharequal}{\kern0pt}\ \isakeywordONE{let}\ \isabound{pi}\ {\isacharequal}{\kern0pt}\ \isafree{l}\ {\isacharbang}{\kern0pt}\ \isavar{i}\ \isakeywordONE{in}\ \isakeywordONE{if}\ \isabound{pi}\ {\isacharequal}{\kern0pt}\ \isavar{i}\ \isakeywordONE{then}\ \isavar{i}\ \isakeywordONE{else}\ \isaconst{rep{\isacharunderscore}{\kern0pt}of}\ \isafree{l}\ \isabound{pi}}}
\end{flushleft}
Looking closely at this definition, we see that this function is only well-defined for some inputs \isa{\isafree{l}} and \isa{\isafree{a}}:
for every element \isa{\isafree{a}\ {\isacharless}{\kern0pt}\ {\isacharbar}{\kern0pt}\isafree{l}{\isacharbar}{\kern0pt}}, its parent must be in the list, i.e.\ we must have \isa{\isafree{l}\ {\isacharbang}{\kern0pt}\ \isafree{a}\ {\isacharless}{\kern0pt}\ {\isacharbar}{\kern0pt}\isafree{l}{\isacharbar}{\kern0pt}},
and the parent pointers must be cycle-free in order for the function to terminate.
Functions in Isabelle/HOL must be total, so Isabelle introduces a constant \isa{\isaconst{rep{\isacharunderscore}{\kern0pt}of{\isacharunderscore}{\kern0pt}dom}\ {\isacharcolon}{\kern0pt}{\isacharcolon}{\kern0pt}\ \isatconst{nat}\ \isatconst{list}\ {\isasymtimes}\ \isatconst{nat}\ {\isasymRightarrow}\ \isatconst{bool}}
that characterises the inputs for which \isa{\isaconst{rep{\isacharunderscore}{\kern0pt}of}} terminates.
Then, it adds \isa{\isaconst{rep{\isacharunderscore}{\kern0pt}of{\isacharunderscore}{\kern0pt}dom}\ {\isacharparenleft}{\kern0pt}\isafree{l}{\isacharcomma}{\kern0pt}\ \isafree{a}{\isacharparenright}{\kern0pt}} as a premise to the defining equation of \isa{\isaconst{rep{\isacharunderscore}{\kern0pt}of}}. 
The intuition above is cast into a predicate \isa{\isaconst{ufa{\isacharunderscore}{\kern0pt}invar}} that defines such well-formed lists \isa{\isafree{l}}.
\begin{flushleft}
{\parindent0pt\isa{\isaconst{ufa{\isacharunderscore}{\kern0pt}invar}\ {\isacharcolon}{\kern0pt}{\isacharcolon}{\kern0pt}\ \isatconst{nat}\ \isatconst{list}\ {\isasymRightarrow}\ \isatconst{bool}}\\[\funheadersep]\isa{\isaconst{ufa{\isacharunderscore}{\kern0pt}invar}\ \isafree{l}\ {\isacharequal}{\kern0pt}\ {\isasymforall}\isabound{i}{\isacharless}{\kern0pt}{\isacharbar}{\kern0pt}\isafree{l}{\isacharbar}{\kern0pt}{\isachardot}{\kern0pt}\ \isaconst{rep{\isacharunderscore}{\kern0pt}of{\isacharunderscore}{\kern0pt}dom}\ {\isacharparenleft}{\kern0pt}\isafree{l}{\isacharcomma}{\kern0pt}\ \isabound{i}{\isacharparenright}{\kern0pt}\ {\isasymand}\ \isafree{l}\ {\isacharbang}{\kern0pt}\ \isabound{i}\ {\isacharless}{\kern0pt}\ {\isacharbar}{\kern0pt}\isafree{l}{\isacharbar}{\kern0pt}}}
\end{flushleft}
Building on the formalisation,
we define the \acrfull{adt} \isa{\isatconst{ufa}} as the set of all \isa{\isafree{l}{\isacharcolon}{\kern0pt}{\isacharcolon}{\kern0pt}\isatconst{nat}\ \isatconst{list}}
that satisfy \isa{\isaconst{ufa{\isacharunderscore}{\kern0pt}invar}\ \isafree{l}}.
\begin{flushleft}
\isakeywordONE{typedef}~\isa{\isatconst{ufa}}~$=$~\isa{{\isacharbraceleft}{\kern0pt}\isabound{l}\ {\isacharbar}{\kern0pt}\ \isaconst{ufa{\isacharunderscore}{\kern0pt}invar}\ \isabound{l}{\isacharbraceright}{\kern0pt}}.
\end{flushleft}
This introduces a new type without any predefined operations.
To equip it with functionality,
we lift the operations on the underlying list due to \citeauthor{uf_isabelle}~\cite{uf_isabelle}
to the \acrshort{adt} using Isabelle's lifting infrastructure~\cite{lifting_transfer},
yielding
\begin{enumerate*}
  \item \isa{\isaconst{ufa{\isacharunderscore}{\kern0pt}{\isasymalpha}}\ {\isacharcolon}{\kern0pt}{\isacharcolon}{\kern0pt}\ \isatconst{ufa}\ {\isasymRightarrow}\ {\isacharparenleft}{\kern0pt}\isatconst{nat}\ {\isasymtimes}\ \isatconst{nat}{\isacharparenright}{\kern0pt}\ \isatconst{set}},
  \item \isa{\isaconst{ufa{\isacharunderscore}{\kern0pt}rep{\isacharunderscore}{\kern0pt}of}\ {\isacharcolon}{\kern0pt}{\isacharcolon}{\kern0pt}\ \isatconst{ufa}\ {\isasymRightarrow}\ \isatconst{nat}\ {\isasymRightarrow}\ \isatconst{nat}},
  \item \isa{\isaconst{ufa{\isacharunderscore}{\kern0pt}init}\ {\isacharcolon}{\kern0pt}{\isacharcolon}{\kern0pt}\ \isatconst{nat}\ {\isasymRightarrow}\ \isatconst{ufa}}, and
  \item \isa{\isaconst{ufa{\isacharunderscore}{\kern0pt}union}\ {\isacharcolon}{\kern0pt}{\isacharcolon}{\kern0pt}\ \isatconst{ufa}\ {\isasymRightarrow}\ \isatconst{nat}\ {\isasymRightarrow}\ \isatconst{nat}\ {\isasymRightarrow}\ \isatconst{ufa}}.
\end{enumerate*}
Their meaning is the following:
\begin{enumerate}
  \item \isa{\isaconst{ufa{\isacharunderscore}{\kern0pt}{\isasymalpha}}\ \isafree{uf}} is the partial equivalence relation represented by \isa{\isafree{uf}},
  \item \isa{\isaconst{ufa{\isacharunderscore}{\kern0pt}rep{\isacharunderscore}{\kern0pt}of}\ \isafree{uf}\ \isafree{x}} is the representative of the equivalence class containing \isa{\isafree{x}},
  \item \isa{\isaconst{ufa{\isacharunderscore}{\kern0pt}init}\ \isafree{n}} initialises the data structure with \isa{\isafree{n}} elements
    with each element being its own representative, and
  \item \isa{\isaconst{ufa{\isacharunderscore}{\kern0pt}union}\ \isafree{uf}\ \isafree{x}\ \isafree{y}} returns a \acrshort{uf} data structure
    where the equivalence classes of \isa{\isafree{x}} and \isa{\isafree{y}} are merged.
    This is implemented by updating the underlying list at index \isa{\isaconst{rep{\isacharunderscore}{\kern0pt}of}\ \isafree{l}\ \isafree{x}} to \isa{\isaconst{rep{\isacharunderscore}{\kern0pt}of}\ \isafree{l}\ \isafree{y}}.
\end{enumerate}
Formally, the above operations fulfil the properties stated below:
\begin{itemize}
  \item \isa{\isaconst{ufa{\isacharunderscore}{\kern0pt}rep{\isacharunderscore}{\kern0pt}of}\ \isavar{uf}\ \isavar{x}\ {\isacharequal}{\kern0pt}\ \isaconst{ufa{\isacharunderscore}{\kern0pt}rep{\isacharunderscore}{\kern0pt}of}\ \isavar{uf}\ \isavar{y}\ \ {\isasymlongleftrightarrow}\ \ {\isacharparenleft}{\kern0pt}\isavar{x}{\isacharcomma}{\kern0pt}\ \isavar{y}{\isacharparenright}{\kern0pt}\ {\isasymin}\ \isaconst{ufa{\isacharunderscore}{\kern0pt}{\isasymalpha}}\ \isavar{uf}} if \isa{{\isacharbraceleft}{\kern0pt}\isafree{x}{\isacharcomma}{\kern0pt}\ \isafree{y}{\isacharbraceright}{\kern0pt}\ {\isasymsubseteq}\ \isaconst{Field}\ {\isacharparenleft}{\kern0pt}\isaconst{ufa{\isacharunderscore}{\kern0pt}{\isasymalpha}}\ \isafree{uf}{\isacharparenright}{\kern0pt}},
  \item \isa{\isaconst{ufa{\isacharunderscore}{\kern0pt}{\isasymalpha}}\ {\isacharparenleft}{\kern0pt}\isaconst{ufa{\isacharunderscore}{\kern0pt}init}\ \isavar{n}{\isacharparenright}{\kern0pt}\ {\isacharequal}{\kern0pt}\ {\isacharbraceleft}{\kern0pt}{\isacharparenleft}{\kern0pt}\isabound{x}{\isacharcomma}{\kern0pt}\ \isabound{x}{\isacharparenright}{\kern0pt}\ {\isacharbar}{\kern0pt}\ \isabound{x}\ {\isacharless}{\kern0pt}\ \isavar{n}{\isacharbraceright}{\kern0pt}}, and
  \item \isa{\isaconst{ufa{\isacharunderscore}{\kern0pt}{\isasymalpha}}\ {\isacharparenleft}{\kern0pt}\isaconst{ufa{\isacharunderscore}{\kern0pt}union}\ \isavar{uf}\ \isavar{x}\ \isavar{y}{\isacharparenright}{\kern0pt}\ {\isacharequal}{\kern0pt}\ \isaconst{per{\isacharunderscore}{\kern0pt}union}\ {\isacharparenleft}{\kern0pt}\isaconst{ufa{\isacharunderscore}{\kern0pt}{\isasymalpha}}\ \isavar{uf}{\isacharparenright}{\kern0pt}\ \isavar{x}\ \isavar{y}}
\end{itemize}
where \isa{\isaconst{per{\isacharunderscore}{\kern0pt}union}\ \isafree{R}\ \isafree{x}\ \isafree{y}} is the equivalence relation
that results from merging the respective equivalence classes in the relation \isa{\isafree{R}} that \isa{\isafree{x}} and \isa{\isafree{y}} belong to.

But what happens if \isa{\isafree{x}} or \isa{\isafree{y}} is not an element of the partial equivalence relation \isa{\isafree{R}}?
In that case, the equivalence relation is unchanged, which means that \isa{\isaconst{per{\isacharunderscore}{\kern0pt}union}\ \isafree{R}\ \isafree{x}\ \isafree{y}\ {\isacharequal}{\kern0pt}\ \isafree{R}}.
This, however, can be seen as a misuse of the \acrshort{uf} data structure,
since we initialise it with a fixed set of elements \isa{\isafree{A}}
and expect the user to only work with these elements.
Therefore, we introduce the following definitions that characterise valid union(s) with regard to this initial set.
\begin{flushleft}
{\parindent0pt\isa{\isaconst{valid{\isacharunderscore}{\kern0pt}union}\ {\isacharcolon}{\kern0pt}{\isacharcolon}{\kern0pt}\ \isatconst{ufa}\ {\isasymRightarrow}\ \isatconst{nat}\ {\isasymRightarrow}\ \isatconst{nat}\ {\isasymRightarrow}\ \isatconst{bool}}\\[\funheadersep]\isa{\isaconst{valid{\isacharunderscore}{\kern0pt}union}\ \isavar{uf}\ \isavar{a}\ \isavar{b}\ {\isacharequal}{\kern0pt}\ \isavar{a}\ {\isasymin}\ \isaconst{Field}\ {\isacharparenleft}{\kern0pt}\isaconst{ufa{\isacharunderscore}{\kern0pt}{\isasymalpha}}\ \isavar{uf}{\isacharparenright}{\kern0pt}\ {\isasymand}\ \isavar{b}\ {\isasymin}\ \isaconst{Field}\ {\isacharparenleft}{\kern0pt}\isaconst{ufa{\isacharunderscore}{\kern0pt}{\isasymalpha}}\ \isavar{uf}{\isacharparenright}{\kern0pt}}} \\[0.75em]
{\parindent0pt\isa{\isaconst{valid{\isacharunderscore}{\kern0pt}unions}\ {\isacharcolon}{\kern0pt}{\isacharcolon}{\kern0pt}\ \isatconst{ufa}\ {\isasymRightarrow}\ {\isacharparenleft}{\kern0pt}\isatconst{nat}\ {\isasymtimes}\ \isatconst{nat}{\isacharparenright}{\kern0pt}\ \isatconst{list}\ {\isasymRightarrow}\ \isatconst{bool}}\\[\funheadersep]\isa{\isaconst{valid{\isacharunderscore}{\kern0pt}unions}\ \isavar{uf}\ \isavar{us}\ {\isacharequal}{\kern0pt}\ {\isasymforall}{\isacharparenleft}{\kern0pt}\isabound{x}{\isacharcomma}{\kern0pt}\ \isabound{y}{\isacharparenright}{\kern0pt}{\isasymin}\isaconst{set}\ \isavar{us}{\isachardot}{\kern0pt}\ \isaconst{valid{\isacharunderscore}{\kern0pt}union}\ \isavar{uf}\ \isabound{x}\ \isabound{y}}}
\end{flushleft}%
\end{isamarkuptext}\isamarkuptrue%
\isadelimdocument
\endisadelimdocument
\isatagdocument
\isamarkupsection{Simple Certifying Union-Find Algorithm\label{sec:ufe_simple}%
}
\isamarkuptrue%
\endisatagdocument
{\isafolddocument}%
\isadelimdocument
\endisadelimdocument
\begin{isamarkuptext}%
Building on the \acrshort{uf} \acrshort{adt}, we now develop a simple \opexplain{} operation that,
for a given list of equations \isa{\isafree{us}{\isacharcolon}{\kern0pt}{\isacharcolon}{\kern0pt}{\isacharparenleft}{\kern0pt}\isatfree{{\isacharprime}{\kern0pt}a}\ {\isasymtimes}\ \isatfree{{\isacharprime}{\kern0pt}a}{\isacharparenright}{\kern0pt}\ \isatconst{list}}, takes two elements \isa{\isafree{x}} and \isa{\isafree{y}}
and produces a certificate that \isa{\isafree{x}\ {\isacharequal}{\kern0pt}\ \isafree{y}} modulo \isa{\isafree{us}}.
The certificate is given in terms of a data type \isa{eq{\isacharunderscore}{\kern0pt}prf}
with its corresponding system \isa{{\isasymturnstile}\isactrlsub {\isacharequal}{\kern0pt}} of inference rules as seen in \cref{fig:eq_prf}.
As expected, we have inference rules that utilise the reflexivity, symmetry, and transitivity of equality as well as an assumption rule.
To improve readability, we use the infix operator $\bigtriangledown$ to denote the proof term for transitivity.
\begin{figure}[b]
  \begin{gather*}
    \inferrule{\isa{\isavar{i}\ {\isacharless}{\kern0pt}\ {\isacharbar}{\kern0pt}\isafree{us}{\isacharbar}{\kern0pt}} \\
       \isa{\isafree{us}\ {\isacharbang}{\kern0pt}\ \isavar{i}\ {\isacharequal}{\kern0pt}\ {\isacharparenleft}{\kern0pt}\isavar{x}{\isacharcomma}{\kern0pt}\ \isavar{y}{\isacharparenright}{\kern0pt}}}{\isa{\isafree{us}\ {\isasymturnstile}\isactrlsub {\isacharequal}{\kern0pt}\ \isaconst{AssmP}\ \isavar{i}\ {\isacharcolon}{\kern0pt}\ {\isacharparenleft}{\kern0pt}\isavar{x}{\isacharcomma}{\kern0pt}\ \isavar{y}{\isacharparenright}{\kern0pt}}} \qquad
    \isa{\mbox{}\inferrule{\mbox{}}{\mbox{\isafree{us}\ {\isasymturnstile}\isactrlsub {\isacharequal}{\kern0pt}\ \isaconst{ReflP}\ \isavar{x}\ {\isacharcolon}{\kern0pt}\ {\isacharparenleft}{\kern0pt}\isavar{x}{\isacharcomma}{\kern0pt}\ \isavar{x}{\isacharparenright}{\kern0pt}}}} \\
    \isa{\mbox{}\inferrule{\mbox{\isafree{us}\ {\isasymturnstile}\isactrlsub {\isacharequal}{\kern0pt}\ \isavar{p}\ {\isacharcolon}{\kern0pt}\ {\isacharparenleft}{\kern0pt}\isavar{x}{\isacharcomma}{\kern0pt}\ \isavar{y}{\isacharparenright}{\kern0pt}}}{\mbox{\isafree{us}\ {\isasymturnstile}\isactrlsub {\isacharequal}{\kern0pt}\ \isaconst{SymP}\ \isavar{p}\ {\isacharcolon}{\kern0pt}\ {\isacharparenleft}{\kern0pt}\isavar{y}{\isacharcomma}{\kern0pt}\ \isavar{x}{\isacharparenright}{\kern0pt}}}} \qquad
    \inferrule{\isa{\isafree{us}\ {\isasymturnstile}\isactrlsub {\isacharequal}{\kern0pt}\ \isafree{p\isactrlsub {\isadigit{1}}}\ {\isacharcolon}{\kern0pt}\ {\isacharparenleft}{\kern0pt}\isavar{x}{\isacharcomma}{\kern0pt}\ \isavar{y}{\isacharparenright}{\kern0pt}} \\
       \isa{\isafree{us}\ {\isasymturnstile}\isactrlsub {\isacharequal}{\kern0pt}\ \isafree{p\isactrlsub {\isadigit{2}}}\ {\isacharcolon}{\kern0pt}\ {\isacharparenleft}{\kern0pt}\isavar{y}{\isacharcomma}{\kern0pt}\ \isavar{z}{\isacharparenright}{\kern0pt}}}{\isa{\isafree{us}\ {\isasymturnstile}\isactrlsub {\isacharequal}{\kern0pt}\ \isafree{p\isactrlsub {\isadigit{1}}}\ \ensuremath{\bigtriangledown}\ \isafree{p\isactrlsub {\isadigit{2}}}\ {\isacharcolon}{\kern0pt}\ {\isacharparenleft}{\kern0pt}\isavar{x}{\isacharcomma}{\kern0pt}\ \isavar{z}{\isacharparenright}{\kern0pt}}}
  \end{gather*}
  \caption{%
    The system of inference rules \isa{{\isasymturnstile}\isactrlsub {\isacharequal}{\kern0pt}} on the data type \isa{eq{\isacharunderscore}{\kern0pt}prf} of proofs.\label{fig:eq_prf}
    We write \isa{\isafree{us}\ {\isasymturnstile}\isactrlsub {\isacharequal}{\kern0pt}\ \isafree{p}\ {\isacharcolon}{\kern0pt}\ {\isacharparenleft}{\kern0pt}\isafree{x}{\isacharcomma}{\kern0pt}\ \isafree{y}{\isacharparenright}{\kern0pt}} to say that \isa{\isafree{p}} proves \isa{\isafree{x}\ {\isacharequal}{\kern0pt}\ \isafree{y}} assuming the equalities \isa{\isafree{us}}.
  }
\end{figure}

We prove that \isa{{\isasymturnstile}\isactrlsub {\isacharequal}{\kern0pt}} is sound and complete with respect to the equivalence relation induced by \isa{\isafree{us}},
i.e.\ the equivalence closure of \isa{\isafree{us}}.
In Isabelle, we define
\begin{flushleft}
\begin{minipage}{0.485\linewidth}
{\parindent0pt\isa{\isaconst{symcl}\ {\isacharcolon}{\kern0pt}{\isacharcolon}{\kern0pt}\ \isatfree{{\isacharprime}{\kern0pt}a}\ \isatconst{rel}\ {\isasymRightarrow}\ \isatfree{{\isacharprime}{\kern0pt}a}\ \isatconst{rel}}\\[\funheadersep]\isa{\isaconst{symcl}\ \isavar{r}\ {\isacharequal}{\kern0pt}\ \isavar{r}\ {\isasymunion}\ \isavar{r}{\isasyminverse}}}
\end{minipage}
\hfill
\begin{minipage}{0.485\linewidth}
{\parindent0pt\isa{\isaconst{equivcl}\ {\isacharcolon}{\kern0pt}{\isacharcolon}{\kern0pt}\ \isatfree{{\isacharprime}{\kern0pt}a}\ \isatconst{rel}\ {\isasymRightarrow}\ \isatfree{{\isacharprime}{\kern0pt}a}\ \isatconst{rel}}\\[\funheadersep]\isa{\isaconst{equivcl}\ \isavar{r}\ {\isacharequal}{\kern0pt}\ {\isacharparenleft}{\kern0pt}\isaconst{symcl}\ \isavar{r}{\isacharparenright}{\kern0pt}\isactrlsup {\isacharasterisk}{\kern0pt}}}
\end{minipage}
\end{flushleft}
and prove the theorem below.
\begin{theorem}[Soundness and Completeness of \isa{{\isasymturnstile}\isactrlsub {\isacharequal}{\kern0pt}}]
\isa{{\normalsize{}If\,}\ \isafree{us}\ {\isasymturnstile}\isactrlsub {\isacharequal}{\kern0pt}\ \isavar{p}\ {\isacharcolon}{\kern0pt}\ {\isacharparenleft}{\kern0pt}\isavar{x}{\isacharcomma}{\kern0pt}\ \isavar{y}{\isacharparenright}{\kern0pt}\ {\normalsize \,then\,}\ {\isacharparenleft}{\kern0pt}\isavar{x}{\isacharcomma}{\kern0pt}\ \isavar{y}{\isacharparenright}{\kern0pt}\ {\isasymin}\ \isaconst{equivcl}\ {\isacharparenleft}{\kern0pt}\isaconst{set}\ \isafree{us}{\isacharparenright}{\kern0pt}{\isachardot}{\kern0pt}}
Conversely, \isa{{\normalsize{}If\,}\ {\isacharparenleft}{\kern0pt}\isavar{x}{\isacharcomma}{\kern0pt}\ \isavar{y}{\isacharparenright}{\kern0pt}\ {\isasymin}\ \isaconst{equivcl}\ {\isacharparenleft}{\kern0pt}\isaconst{set}\ \isafree{us}{\isacharparenright}{\kern0pt}\ {\normalsize \,then\,}\ {\isasymexists}\isabound{p}{\isachardot}{\kern0pt}\ \isafree{us}\ {\isasymturnstile}\isactrlsub {\isacharequal}{\kern0pt}\ \isabound{p}\ {\isacharcolon}{\kern0pt}\ {\isacharparenleft}{\kern0pt}\isavar{x}{\isacharcomma}{\kern0pt}\ \isavar{y}{\isacharparenright}{\kern0pt}{\isachardot}{\kern0pt}}
\end{theorem}

Our goal is to implement the \opexplain{} operation using a \acrshort{uf} data structure,
so we fix an initial \acrshort{uf} data structure \isa{\isafree{uf}}.
For a list of equations \isa{\isafree{us}} or, in terms of the \acrshort{uf} data structure, \opunion{} operations,
the current state of the \acrshort{uf} data structure is then equal to \isa{\isaconst{ufa{\isacharunderscore}{\kern0pt}unions}\ \isafree{uf}\ \isafree{us}}
where we define
\begin{flushleft}
{\parindent0pt\isa{\isaconst{ufa{\isacharunderscore}{\kern0pt}unions}\ {\isacharcolon}{\kern0pt}{\isacharcolon}{\kern0pt}\ \isatconst{ufa}\ {\isasymRightarrow}\ {\isacharparenleft}{\kern0pt}\isatconst{nat}\ {\isasymtimes}\ \isatconst{nat}{\isacharparenright}{\kern0pt}\ \isatconst{list}\ {\isasymRightarrow}\ \isatconst{ufa}}\\[\funheadersep]\isa{\isaconst{ufa{\isacharunderscore}{\kern0pt}unions}\ {\isacharequal}{\kern0pt}\ \isaconst{foldl}\ {\isacharparenleft}{\kern0pt}{\isasymlambda}\isabound{uf}\ {\isacharparenleft}{\kern0pt}\isabound{x}{\isacharcomma}{\kern0pt}\ \isabound{y}{\isacharparenright}{\kern0pt}{\isachardot}{\kern0pt}\ \isaconst{ufa{\isacharunderscore}{\kern0pt}union}\ \isabound{uf}\ \isabound{x}\ \isabound{y}{\isacharparenright}{\kern0pt}}}.
\end{flushleft}
Here, we require the unions \isa{\isafree{us}} to be valid unions with respect to \isa{\isafree{uf}}.
Moreover, it must hold that \isa{\isaconst{ufa{\isacharunderscore}{\kern0pt}{\isasymalpha}}\ \isafree{uf}\ {\isasymsubseteq}\ \isaconst{Id}}
because the only way to justify an equality from an empty list of equations using \isa{{\isasymturnstile}\isactrlsub {\isacharequal}{\kern0pt}} is by reflexivity.
Finally, we also constrain \isa{\isafree{us}} to be \emph{effective} unions
meaning that no union shall be redundant with respect to the unions preceeding it.
Note that redundant unions have no effect on the state of the \acrshort{uf} data structure anyways
so there is no need to record them.
We formalise effectiveness with the following definitions.
\begin{flushleft}
{\parindent0pt\isa{\isaconst{eff{\isacharunderscore}{\kern0pt}union}\ {\isacharcolon}{\kern0pt}{\isacharcolon}{\kern0pt}\ \isatconst{ufa}\ {\isasymRightarrow}\ \isatconst{nat}\ {\isasymRightarrow}\ \isatconst{nat}\ {\isasymRightarrow}\ \isatconst{bool}}\\[\funheadersep]\isa{\isaconst{eff{\isacharunderscore}{\kern0pt}union}\ \isavar{uf}\ \isavar{a}\ \isavar{b}\ {\isacharequal}{\kern0pt}\ \isaconst{valid{\isacharunderscore}{\kern0pt}union}\ \isavar{uf}\ \isavar{a}\ \isavar{b}\ {\isasymand}\ \isaconst{ufa{\isacharunderscore}{\kern0pt}rep{\isacharunderscore}{\kern0pt}of}\ \isavar{uf}\ \isavar{a}\ {\isasymnoteq}\ \isaconst{ufa{\isacharunderscore}{\kern0pt}rep{\isacharunderscore}{\kern0pt}of}\ \isavar{uf}\ \isavar{b}}} \\[0.75em]
{\parindent0pt\isa{\isaconst{eff{\isacharunderscore}{\kern0pt}unions}\ {\isacharcolon}{\kern0pt}{\isacharcolon}{\kern0pt}\ \isatconst{ufa}\ {\isasymRightarrow}\ {\isacharparenleft}{\kern0pt}\isatconst{nat}\ {\isasymtimes}\ \isatconst{nat}{\isacharparenright}{\kern0pt}\ \isatconst{list}\ {\isasymRightarrow}\ \isatconst{bool}}\\[\funheadersep]\isa{\isaconst{eff{\isacharunderscore}{\kern0pt}unions}\ \isavar{uf}\ {\isacharbrackleft}{\kern0pt}{\isacharbrackright}{\kern0pt}\ {\isacharequal}{\kern0pt}\ \isaconst{True}\isanewline
\isaconst{eff{\isacharunderscore}{\kern0pt}unions}\ \isavar{uf}\ {\isacharparenleft}{\kern0pt}{\isacharparenleft}{\kern0pt}\isavar{a}{\isacharcomma}{\kern0pt}\ \isavar{b}{\isacharparenright}{\kern0pt}\ {\isacharhash}{\kern0pt}\ \isavar{us}{\isacharparenright}{\kern0pt}\ {\isacharequal}{\kern0pt}\isanewline
\isaindent{\ \ }\ \isaconst{eff{\isacharunderscore}{\kern0pt}union}\ \isavar{uf}\ \isavar{a}\ \isavar{b}\ {\isasymand}\ \isaconst{eff{\isacharunderscore}{\kern0pt}unions}\ {\isacharparenleft}{\kern0pt}\isaconst{ufa{\isacharunderscore}{\kern0pt}union}\ \isavar{uf}\ \isavar{a}\ \isavar{b}{\isacharparenright}{\kern0pt}\ \isavar{us}}}
\end{flushleft}
Similarly to \isa{\isatconst{ufa}}, we encapsulate pairs \isa{{\isacharparenleft}{\kern0pt}\isafree{uf}{\isacharcomma}{\kern0pt}\ \isafree{us}{\isacharparenright}{\kern0pt}}
that are well-formed with respect to the constraints above by an \acrshort{adt} \isa{\isatconst{ufe}}.
We choose this simple representation of the \acrshort{ufe} data structure to ease formal reasoning,
while a more efficient implementation is described in \cref{sec:imperative_hol}.

\begin{flushleft}
\isakeywordONE{typedef}~\isa{\isatconst{ufe}}~$=$~\isa{{\isacharbraceleft}{\kern0pt}{\isacharparenleft}{\kern0pt}\isabound{uf}{\isacharcomma}{\kern0pt}\ \isabound{us}{\isacharparenright}{\kern0pt}\ {\isacharbar}{\kern0pt}\ \isaconst{ufa{\isacharunderscore}{\kern0pt}{\isasymalpha}}\ \isabound{uf}\ {\isasymsubseteq}\ \isaconst{Id}\ {\isasymand}\ \isaconst{eff{\isacharunderscore}{\kern0pt}unions}\ \isabound{uf}\ \isabound{us}{\isacharbraceright}{\kern0pt}}
\end{flushleft}
We lift operations on such pairs \isa{{\isacharparenleft}{\kern0pt}\isafree{uf}{\isacharcomma}{\kern0pt}\ \isafree{us}{\isacharparenright}{\kern0pt}} to obtain
\begin{enumerate*}
  \item \isa{\isaconst{unions}\ {\isacharcolon}{\kern0pt}{\isacharcolon}{\kern0pt}\ \isatconst{ufe}\ {\isasymRightarrow}\ {\isacharparenleft}{\kern0pt}\isatconst{nat}\ {\isasymtimes}\ \isatconst{nat}{\isacharparenright}{\kern0pt}\ \isatconst{list}},
  \item \isa{\isaconst{uf{\isacharunderscore}{\kern0pt}ds}\ {\isacharcolon}{\kern0pt}{\isacharcolon}{\kern0pt}\ \isatconst{ufe}\ {\isasymRightarrow}\ \isatconst{ufa}},
  \item \isa{\isaconst{ufe{\isacharunderscore}{\kern0pt}init}\ {\isacharcolon}{\kern0pt}{\isacharcolon}{\kern0pt}\ \isatconst{nat}\ {\isasymRightarrow}\ \isatconst{ufe}}, and
  \item both \isa{\isaconst{ufe{\isacharunderscore}{\kern0pt}union}\ {\isacharcolon}{\kern0pt}{\isacharcolon}{\kern0pt}\ \isatconst{ufe}\ {\isasymRightarrow}\ \isatconst{nat}\ {\isasymRightarrow}\ \isatconst{nat}\ {\isasymRightarrow}\ \isatconst{ufe}} and its dual
  \item \isa{\isaconst{rollback}\ {\isacharcolon}{\kern0pt}{\isacharcolon}{\kern0pt}\ \isatconst{ufe}\ {\isasymRightarrow}\ \isatconst{ufe}}.
\end{enumerate*}
The meaning of these operations is the following:
\begin{enumerate*}
  \item \isa{\isaconst{unions}\ \isafree{ufe}} is the list of unions \isa{\isafree{us}},
  \item \isa{\isaconst{uf{\isacharunderscore}{\kern0pt}ds}\ \isafree{ufe}} represents the current state of the \acrshort{uf} data structure, i.e.\ \isa{\isaconst{ufa{\isacharunderscore}{\kern0pt}unions}\ \isafree{uf}\ \isafree{us}},
  \item \isa{\isaconst{ufe{\isacharunderscore}{\kern0pt}init}\ \isafree{n}} initialises the data structure with \isa{\isafree{n}} elements and an empty list of unions,
  \item \isa{\isaconst{ufe{\isacharunderscore}{\kern0pt}union}\ \isafree{ufe}\ \isafree{a}\ \isafree{b}} appends an effective union \isa{{\isacharparenleft}{\kern0pt}\isafree{a}{\isacharcomma}{\kern0pt}\ \isafree{b}{\isacharparenright}{\kern0pt}} to \isa{\isafree{us}}, and
  \item \isa{\isaconst{rollback}\ \isafree{ufe}} removes the last union from \isa{\isafree{us}}.
\end{enumerate*}
Furthermore, we lift the remaining operations on \isa{\isatconst{ufa}} to \isa{\isatconst{ufe}} via \isa{\isaconst{uf{\isacharunderscore}{\kern0pt}ds}},
replacing the prefix \textsf{ufa} by \textsf{ufe}.
For example, we lift \isa{\isaconst{ufa{\isacharunderscore}{\kern0pt}rep{\isacharunderscore}{\kern0pt}of}} by letting \isa{\isaconst{ufe{\isacharunderscore}{\kern0pt}rep{\isacharunderscore}{\kern0pt}of}\ \isafree{ufe}\ {\isasymequiv}\ \isaconst{ufa{\isacharunderscore}{\kern0pt}rep{\isacharunderscore}{\kern0pt}of}\ {\isacharparenleft}{\kern0pt}\isaconst{uf{\isacharunderscore}{\kern0pt}ds}\ \isafree{ufe}{\isacharparenright}{\kern0pt}}.

\begin{figure*}
\begin{flushleft}
{\parindent0pt\isa{\isaconst{explain}\ {\isacharcolon}{\kern0pt}{\isacharcolon}{\kern0pt}\ \isatconst{ufe}\ {\isasymRightarrow}\ \isatconst{nat}\ {\isasymRightarrow}\ \isatconst{nat}\ {\isasymRightarrow}\ \isatconst{nat}\ \isatconst{eq{\isacharunderscore}{\kern0pt}prf}}\\[\funheadersep]\isa{\isaconst{explain}\ \isavar{ufe}\ \isavar{x}\ \isavar{y}\ {\isacharequal}{\kern0pt}\isanewline
\isaindent{\ \ }\ \isakeywordONE{if}\ \isaconst{unions}\ \isavar{ufe}\ {\isacharequal}{\kern0pt}\ {\isacharbrackleft}{\kern0pt}{\isacharbrackright}{\kern0pt}\ \isakeywordONE{then}\ \isaconst{ReflP}\ \isavar{x}\isanewline
\isaindent{\ \ \ }\isakeywordONE{else}\ \isakeywordONE{let}\ \isabound{ufe{\isadigit{0}}}\ {\isacharequal}{\kern0pt}\ \isaconst{rollback}\ \isavar{ufe}{\isacharsemicolon}{\kern0pt}\ {\isacharparenleft}{\kern0pt}\isabound{a}{\isacharcomma}{\kern0pt}\ \isabound{b}{\isacharparenright}{\kern0pt}\ {\isacharequal}{\kern0pt}\ \isaconst{last}\ {\isacharparenleft}{\kern0pt}\isaconst{unions}\ \isavar{ufe}{\isacharparenright}{\kern0pt}{\isacharsemicolon}{\kern0pt}\isanewline
\isaindent{\ \ \ \isakeywordONE{else}\ \isakeywordONE{let}\ }\isabound{a{\isacharunderscore}{\kern0pt}b{\isacharunderscore}{\kern0pt}P}\ {\isacharequal}{\kern0pt}\ \isaconst{AssmP}\ {\isacharbar}{\kern0pt}\isaconst{unions}\ \isabound{ufe{\isadigit{0}}}{\isacharbar}{\kern0pt}\isanewline
\isaindent{\ \ \ \isakeywordONE{else}\ }\isakeywordONE{in}\ \isakeywordONE{if}\ \isaconst{ufe{\isacharunderscore}{\kern0pt}rep{\isacharunderscore}{\kern0pt}of}\ \isabound{ufe{\isadigit{0}}}\ \isavar{x}\ {\isacharequal}{\kern0pt}\ \isaconst{ufe{\isacharunderscore}{\kern0pt}rep{\isacharunderscore}{\kern0pt}of}\ \isabound{ufe{\isadigit{0}}}\ \isavar{y}\ \isakeywordONE{then}\ \isaconst{explain}\ \isabound{ufe{\isadigit{0}}}\ \isavar{x}\ \isavar{y}\isanewline
\isaindent{\ \ \ \isakeywordONE{else}\ \isakeywordONE{in}\ }\isakeywordONE{else}\ \isakeywordONE{if}\ \isaconst{ufe{\isacharunderscore}{\kern0pt}rep{\isacharunderscore}{\kern0pt}of}\ \isabound{ufe{\isadigit{0}}}\ \isavar{x}\ {\isacharequal}{\kern0pt}\ \isaconst{ufe{\isacharunderscore}{\kern0pt}rep{\isacharunderscore}{\kern0pt}of}\ \isabound{ufe{\isadigit{0}}}\ \isabound{a}\isanewline
\isaindent{\ \ \ \isakeywordONE{else}\ \isakeywordONE{in}\ \isakeywordONE{else}\ }\isakeywordONE{then}\ \isaconst{explain}\ \isabound{ufe{\isadigit{0}}}\ \isavar{x}\ \isabound{a}\ \ensuremath{\bigtriangledown}\ \isabound{a{\isacharunderscore}{\kern0pt}b{\isacharunderscore}{\kern0pt}P}\ \ensuremath{\bigtriangledown}\ \isaconst{explain}\ \isabound{ufe{\isadigit{0}}}\ \isabound{b}\ \isavar{y}\isanewline
\isaindent{\ \ \ \isakeywordONE{else}\ \isakeywordONE{in}\ \isakeywordONE{else}\ }\isakeywordONE{else}\ \isaconst{explain}\ \isabound{ufe{\isadigit{0}}}\ \isavar{x}\ \isabound{b}\ \ensuremath{\bigtriangledown}\ \isaconst{SymP}\ \isabound{a{\isacharunderscore}{\kern0pt}b{\isacharunderscore}{\kern0pt}P}\ \ensuremath{\bigtriangledown}\ \isaconst{explain}\ \isabound{ufe{\isadigit{0}}}\ \isabound{a}\ \isavar{y}}} \\[0.75em]
{\parindent0pt\isa{\isaconst{explain{\isacharunderscore}{\kern0pt}partial}\ {\isacharcolon}{\kern0pt}{\isacharcolon}{\kern0pt}\ \isatconst{ufe}\ {\isasymRightarrow}\ \isatconst{nat}\ {\isasymRightarrow}\ \isatconst{nat}\ {\isasymRightarrow}\ \isatconst{nat}\ \isatconst{eq{\isacharunderscore}{\kern0pt}prf}\ \isatconst{option}}\\[\funheadersep]\isa{\isaconst{explain{\isacharunderscore}{\kern0pt}partial}\ \isavar{ufe}\ \isavar{x}\ \isavar{y}\ {\isacharequal}{\kern0pt}\isanewline
\isaindent{\ \ }\ \isakeywordONE{if}\ {\isacharparenleft}{\kern0pt}\isavar{x}{\isacharcomma}{\kern0pt}\ \isavar{y}{\isacharparenright}{\kern0pt}\ {\isasymin}\ \isaconst{equivcl}\ {\isacharparenleft}{\kern0pt}\isaconst{set}\ {\isacharparenleft}{\kern0pt}\isaconst{unions}\ \isavar{ufe}{\isacharparenright}{\kern0pt}{\isacharparenright}{\kern0pt}\ \isakeywordONE{then}\ \isaconst{Some}\ {\isacharparenleft}{\kern0pt}\isaconst{explain}\ \isavar{ufe}\ \isavar{x}\ \isavar{y}{\isacharparenright}{\kern0pt}\isanewline
\isaindent{\ \ \ }\isakeywordONE{else}\ \isaconst{None}}}
\end{flushleft}
  \caption{A simple implementation of the \opexplain{} operation.\label{fig:explain}
  }
\end{figure*}

At last, we implement the \opexplain{} operation as depicted in \cref{fig:explain}.
The algorithm assumes that the given elements \isa{\isafree{x}} and \isa{\isafree{y}} are equal modulo \isa{\isaconst{unions}\ \isafree{ufe}}.

If \isa{\isaconst{unions}\ \isafree{ufe}\ {\isacharequal}{\kern0pt}\ {\isacharbrackleft}{\kern0pt}{\isacharbrackright}{\kern0pt}}, then \isa{\isafree{x}} and \isa{\isafree{y}} must be equal which we certify with \isa{\isaconst{ReflP}\ \isafree{x}}.

Otherwise, we distinguish between two cases:
\begin{enumerate*}
  \item The elements \isa{\isafree{x}} and \isa{\isafree{y}} are already equal modulo \isa{\isaconst{unions}\ {\isacharparenleft}{\kern0pt}\isaconst{rollback}\ \isafree{ufe}{\isacharparenright}{\kern0pt}},
    so we proceed recursively with \isa{\isaconst{rollback}\ \isafree{ufe}}.
  \item In the case where the most recent equation \isa{\isafree{a}\ {\isacharequal}{\kern0pt}\ \isafree{b}} is necessary for \isa{\isafree{x}\ {\isacharequal}{\kern0pt}\ \isafree{y}} to hold,
    we either have \isa{\isafree{x}\ {\isacharequal}{\kern0pt}\ \isafree{a}} and \isa{\isafree{b}\ {\isacharequal}{\kern0pt}\ \isafree{y}}
    or \isa{\isafree{x}\ {\isacharequal}{\kern0pt}\ \isafree{b}} and \isa{\isafree{a}\ {\isacharequal}{\kern0pt}\ \isafree{y}} modulo \isa{\isaconst{unions}\ {\isacharparenleft}{\kern0pt}\isaconst{rollback}\ \isafree{ufe}{\isacharparenright}{\kern0pt}}.
    Assuming the former holds --- the other case is symmetric ---
    we recursively construct the certificates for \isa{\isafree{x}\ {\isacharequal}{\kern0pt}\ \isafree{a}} and \isa{\isafree{b}\ {\isacharequal}{\kern0pt}\ \isafree{y}}.
    Together with the assumption \isa{\isafree{a}\ {\isacharequal}{\kern0pt}\ \isafree{b}}, we obtain \isa{\isafree{x}\ {\isacharequal}{\kern0pt}\ \isafree{y}} by transitivity.
\end{enumerate*}
The termination of \isa{\isaconst{explain}} is easily proven
because the length of \isa{\isaconst{unions}\ \isafree{ufe}} decreases in each recursive call.
Dually, this termination argument gives rise to the following induction principle.
\begin{lemma}[Induction on \isa{\isatconst{ufe}}]\label{thm:ufe_induct}
In order to prove \isa{\isavar{P}\ \isavar{ufe}} for all \isa{\isafree{ufe}},
we have two inductive cases, both fixing an arbitrary \isa{\isafree{ufe}}:
\begin{enumerate*}
  \item Assume \isa{\isaconst{ufe{\isacharunderscore}{\kern0pt}{\isasymalpha}}\ \isafree{ufe}\ {\isasymsubseteq}\ \isaconst{Id}} as well as \isa{\isaconst{unions}\ \isafree{ufe}\ {\isacharequal}{\kern0pt}\ {\isacharbrackleft}{\kern0pt}{\isacharbrackright}{\kern0pt}}
    and show \isa{\isafree{P}\ \isafree{ufe}}.
  \item Assume \isa{\isaconst{eff{\isacharunderscore}{\kern0pt}union}\ {\isacharparenleft}{\kern0pt}\isaconst{uf{\isacharunderscore}{\kern0pt}ds}\ \isafree{ufe}{\isacharparenright}{\kern0pt}\ \isafree{a}\ \isafree{b}} as well as \isa{\isafree{P}\ \isafree{ufe}}
    and show \isa{\isafree{P}\ {\isacharparenleft}{\kern0pt}\isaconst{ufe{\isacharunderscore}{\kern0pt}union}\ \isafree{ufe}\ \isafree{a}\ \isafree{b}{\isacharparenright}{\kern0pt}}.
\end{enumerate*}
\end{lemma}

We condense the intuition above into the completeness \lcnamecref{thm:explain_complete} below,
which we prove using the induction principle from \cref{thm:ufe_induct}.
\begin{theorem}[Completeness of \isa{\isaconst{explain}}\label{thm:explain_complete}]
\isa{{\normalsize{}If\,}\ {\isacharparenleft}{\kern0pt}\isavar{x}{\isacharcomma}{\kern0pt}\ \isavar{y}{\isacharparenright}{\kern0pt}\ {\isasymin}\ \isaconst{equivcl}\ {\isacharparenleft}{\kern0pt}\isaconst{set}\ {\isacharparenleft}{\kern0pt}\isaconst{unions}\ \isavar{ufe}{\isacharparenright}{\kern0pt}{\isacharparenright}{\kern0pt}\ {\normalsize \,then\,}\ \isaconst{unions}\ \isavar{ufe}\ {\isasymturnstile}\isactrlsub {\isacharequal}{\kern0pt}\ \isaconst{explain}\ \isavar{ufe}\ \isavar{x}\ \isavar{y}\ {\isacharcolon}{\kern0pt}\ {\isacharparenleft}{\kern0pt}\isavar{x}{\isacharcomma}{\kern0pt}\ \isavar{y}{\isacharparenright}{\kern0pt}{\isachardot}{\kern0pt}}
\end{theorem}

The \isa{\isaconst{explain}} function is not sound, though.
This is because it always returns a certificate, even if \isa{\isafree{x}} and \isa{\isafree{y}} are not equal modulo \isa{\isafree{us}}.
To account for this case, we wrap \isa{\isaconst{explain}} into a partial function \isa{\isaconst{explain{\isacharunderscore}{\kern0pt}partial}} (cf.\ \cref{fig:explain})
that fails if \isa{\isafree{x}\ {\isacharequal}{\kern0pt}\ \isafree{y}} is not provable.
Soundness and completeness can then be lifted from the soundness of \isa{{\isasymturnstile}\isactrlsub {\isacharequal}{\kern0pt}} and the completeness of \isa{\isaconst{explain}}, respectively.
Note that membership of \isa{\isaconst{equivcl}} can actually be implemented using \acrshort{uf} operations as the following \lcnamecref{thm:equivcl_iff} demonstrates.
Moreover, it holds that \isa{\isafree{x}\ {\isasymin}\ \isaconst{Field}\ {\isacharparenleft}{\kern0pt}\isaconst{ufa{\isacharunderscore}{\kern0pt}{\isasymalpha}}\ \isafree{uf}{\isacharparenright}{\kern0pt}\ \ {\isasymlongleftrightarrow}\ \ \isafree{x}\ {\isacharless}{\kern0pt}\ \isafree{n}} where \isa{\isafree{n}} is the length of the list representing \isa{\isafree{uf}}.
\begin{lemma}\label{thm:equivcl_iff}
We have \isa{{\isacharparenleft}{\kern0pt}\isavar{x}{\isacharcomma}{\kern0pt}\ \isavar{y}{\isacharparenright}{\kern0pt}\ {\isasymin}\ \isaconst{equivcl}\ {\isacharparenleft}{\kern0pt}\isaconst{set}\ {\isacharparenleft}{\kern0pt}\isaconst{unions}\ \isavar{ufe}{\isacharparenright}{\kern0pt}{\isacharparenright}{\kern0pt}} \emph{iff}
\isa{\isavar{x}\ {\isacharequal}{\kern0pt}\ \isavar{y}\ {\isasymor}\ \isavar{x}\ {\isasymin}\ \isaconst{Field}\ {\isacharparenleft}{\kern0pt}\isaconst{ufe{\isacharunderscore}{\kern0pt}{\isasymalpha}}\ \isavar{ufe}{\isacharparenright}{\kern0pt}\ {\isasymand}\ \isavar{y}\ {\isasymin}\ \isaconst{Field}\ {\isacharparenleft}{\kern0pt}\isaconst{ufe{\isacharunderscore}{\kern0pt}{\isasymalpha}}\ \isavar{ufe}{\isacharparenright}{\kern0pt}\ {\isasymand}\ \isaconst{ufe{\isacharunderscore}{\kern0pt}rep{\isacharunderscore}{\kern0pt}of}\ \isavar{ufe}\ \isavar{x}\ {\isacharequal}{\kern0pt}\ \isaconst{ufe{\isacharunderscore}{\kern0pt}rep{\isacharunderscore}{\kern0pt}of}\ \isavar{ufe}\ \isavar{y}}.
\end{lemma}%
\end{isamarkuptext}\isamarkuptrue%
\isadelimdocument
\endisadelimdocument
\isatagdocument
\isamarkupsection{Efficient Certifying Union-Find Algorithm\label{sec:ufe_efficient}%
}
\isamarkuptrue%
\endisatagdocument
{\isafolddocument}%
\isadelimdocument
\endisadelimdocument
\begin{isamarkuptext}%
In the previous section, we developed an \opexplain{} operation that iteratively removes the most recent union from a list of unions,
identifying which of them, when viewed as equalities, are necessary to prove the input arguments equal.
Iterating through all equalities seems inefficient, though.
Intuitively, we aim to return only those on the path between the arguments,
viewing the equalities as an undirected graph.
To realise this, \citeauthor{congcl_proofs}~\cite{congcl_proofs} use a \acrshort{uf} data structure
represented as forest of rooted trees as described in \cref{sec:uf_background}.
They modify the data structure such that, for each union between \isa{\isafree{a}} and \isa{\isafree{b}},
the newly added edge in the forest gets annotated with \isa{{\isacharparenleft}{\kern0pt}\isafree{a}{\isacharcomma}{\kern0pt}\ \isafree{b}{\isacharparenright}{\kern0pt}}.
To understand why this allows for a more efficient implementation of the \opexplain{} operation,
suppose that we want to certify that \isa{\isafree{x}} is equal to \isa{\isafree{y}}.
Clearly, only the edges of the subtree rooted at the \acrfull{lca} of \isa{\isafree{x}} and \isa{\isafree{y}},
as illustrated in \cref{fig:explain'_abs}, are relevant to explain why \isa{\isafree{x}} is equal to \isa{\isafree{y}}.
Furthermore, let \isa{{\isacharparenleft}{\kern0pt}\isafree{a}{\isacharcomma}{\kern0pt}\ \isafree{b}{\isacharparenright}{\kern0pt}} be the most recent union on either of the paths from the \acrshort{lca}
to \isa{\isafree{x}} or \isa{\isafree{y}}.
Here, we assume w.l.o.g.\ that \isa{{\isacharparenleft}{\kern0pt}\isafree{a}{\isacharcomma}{\kern0pt}\ \isafree{b}{\isacharparenright}{\kern0pt}} is on the path to \isa{\isafree{x}}.
The corresponding edge separates the tree rooted at the \acrshort{lca} into two subtrees as indicated by the patterns, 
one containing \isa{\isafree{a}} and the other one \isa{\isafree{b}}.
Moreover, the paths from the \acrshort{lca} can't overlap, so \isa{\isafree{x}} and \isa{\isafree{y}} also belong to different subtrees.
Ultimately, to certify the equality of \isa{\isafree{x}} and \isa{\isafree{y}},
we recursively prove that \isa{\isafree{x}} is equal to \isa{\isafree{a}} and \isa{\isafree{y}} to \isa{\isafree{b}}.
Then, we put everything together using transitivity and the equality \isa{\isafree{a}\ {\isacharequal}{\kern0pt}\ \isafree{b}}.
This terminates since \isa{{\isacharparenleft}{\kern0pt}\isafree{a}{\isacharcomma}{\kern0pt}\ \isafree{b}{\isacharparenright}{\kern0pt}} is the most recent union and we only consider less recent unions in the recursive steps.
All in all, this gives a $\mathcal{O}(k \log n)$ \opexplain{} operation on a \acrshort{uf} data structure with union-by-size,
where $k$ is the number of unions required for an explanation~\cite{congcl_proofs}.
This is an improvement over the naive algorithm where we iterate over all (up to $n - 1$) unions.
\begin{figure}
  \centering
  \begin{tikzpicture}[
    >=latex, node distance=0.5cm,
    aside/.append style={pattern=north east lines, pattern color=gray!65},
    bside/.append style={pattern={Dots[radius=0.65pt]}, pattern color=gray!65}
    ]
    \node[draw, circle, preaction={fill, white}, style=bside] (lca) {\acrshort{lca}};
    \begin{pgfonlayer}{background}
      \begin{scope}[shape=isosceles triangle, shape border rotate=90, minimum height=1.3cm, minimum width=1.85cm]
        \node[draw, anchor=north, below left=1.35cm and 3cm of lca.south west, aside] (l) {};
        \path (l.north) -- node[pos=0.54, draw, anchor=north, bside] (m) {} (lca.west);
        \node[draw, anchor=north, yshift=0.25cm, bside] (r) at (lca.south) {};
      \end{scope}
    \end{pgfonlayer}

    \node[draw, circle, above right=0.18cm and 0.05cm of l.south west] (x) {\isa{\isafree{x}}};
    \node[draw, circle, left=0.1cm of l.east] (a) {\isa{\isafree{a}}};
    \node[draw, circle, above left=of r.south east] (y) {\isa{\isafree{y}}};
    \node[draw, circle, above right=of m.south west] (b) {\isa{\isafree{b}}};
    
    \draw[->] (l.north) -- node[above, sloped] {\isa{{\isacharparenleft}{\kern0pt}\isafree{a}{\isacharcomma}{\kern0pt}\ \isafree{b}{\isacharparenright}{\kern0pt}}} (m.north);
    \draw[->, dashed] (m.north) -- (lca.west);
    \draw[solid] (lca.north) -- ++(0,0.1);
    \draw[dashed] (lca.north) ++(0,0.1) -- ++(0,0.4);
  \end{tikzpicture}
  \caption{%
    For arguments \isa{\isafree{x}} and \isa{\isafree{y}},
    \isa{\isaconst{explain{\isacharprime}{\kern0pt}}} considers an edge annotated with \isa{{\isacharparenleft}{\kern0pt}\isafree{a}{\isacharcomma}{\kern0pt}\ \isafree{b}{\isacharparenright}{\kern0pt}} that separates the subtree
    rooted at the \acrshort{lca} of \isa{\isafree{x}} and \isa{\isafree{y}} into two subtrees:
    one containing \isa{\isafree{x}} and \isa{\isafree{a}} and the other one containing \isa{\isafree{y}} and \isa{\isafree{b}}.\label{fig:explain'_abs}
  }
\end{figure}

To achieve optimal almost constant running time for \opunion{} and \opfind{},
we need path compression in addition to union-by-size.
Path compression, however, is incompatible with the \opexplain{} operation,
so \citeauthor{congcl_proofs}~\cite{congcl_proofs} propose to maintain two copies of the \acrshort{uf} data structure,
one with and one without path compression.%
\end{isamarkuptext}\isamarkuptrue%
\isadelimdocument
\endisadelimdocument
\isatagdocument
\isamarkupsubsection{Implementation%
}
\isamarkuptrue%
\endisatagdocument
{\isafolddocument}%
\isadelimdocument
\endisadelimdocument
\begin{isamarkuptext}%
To obtain an efficient \opexplain{} operation,
we leverage the structure of the \acrshort{uf} data structure,
which is a forest of rooted trees.
We make this structure accessible by defining a function \isa{\isaconst{ufa{\isacharunderscore}{\kern0pt}parent{\isacharunderscore}{\kern0pt}of}\ {\isacharcolon}{\kern0pt}{\isacharcolon}{\kern0pt}\ \isatconst{ufa}\ {\isasymRightarrow}\ \isatconst{nat}\ {\isasymRightarrow}\ \isatconst{nat}} via lifting,
where \isa{\isaconst{ufa{\isacharunderscore}{\kern0pt}parent{\isacharunderscore}{\kern0pt}of}\ \isafree{uf}\ \isafree{x}} returns the parent of \isa{\isafree{x}}.
This function is related to \isa{\isaconst{ufa{\isacharunderscore}{\kern0pt}rep{\isacharunderscore}{\kern0pt}of}} in the obvious way, i.e.\ we have
\isa{\isaconst{ufa{\isacharunderscore}{\kern0pt}parent{\isacharunderscore}{\kern0pt}of}\ \isavar{uf}\ \isafree{x}\ {\isacharequal}{\kern0pt}\ \isafree{x}\ \emph{iff}\ \isaconst{ufa{\isacharunderscore}{\kern0pt}rep{\isacharunderscore}{\kern0pt}of}\ \isavar{uf}\ \isafree{x}\ {\isacharequal}{\kern0pt}\ \isafree{x}}
for \isa{\isafree{x}\ {\isasymin}\ \isaconst{Field}\ {\isacharparenleft}{\kern0pt}\isaconst{ufa{\isacharunderscore}{\kern0pt}{\isasymalpha}}\ \isavar{uf}{\isacharparenright}{\kern0pt}}.
With this, we formalise the concept of \acrshort{ufe} forests,
define the notion of associated unions within this forest,
and introduce the two auxiliary functions that are the ingredients to the efficient \opexplain{} operation.%
\end{isamarkuptext}\isamarkuptrue%
\isadelimdocument
\endisadelimdocument
\isatagdocument
\isamarkupsubsubsection{\acrshort{ufe} forests%
}
\isamarkuptrue%
\endisatagdocument
{\isafolddocument}%
\isadelimdocument
\endisadelimdocument
\begin{isamarkuptext}%
It is often useful to view the forest of rooted trees underpinning the \acrshort{uf} data structure as a graph.
For this purpose,
we use the graph theory library~\cite{graph_theory} due to \citeauthor{graph_theory},
which is available as an entry of the \acrshort{afp}~\cite{graph_theory_afp}.
The library allows us to represent a graph as a record with the fields \isa{\isaconst{verts}} and \isa{\isaconst{arcs}}
for its vertices and edges,
where edges are pairs of vertices.
The forest induced by a \acrshort{uf} data structure is then defined as follows.
\begin{flushleft}
{\parindent0pt\isa{\isaconst{ufa{\isacharunderscore}{\kern0pt}forest{\isacharunderscore}{\kern0pt}of}\ \isavar{uf}\ {\isacharequal}{\kern0pt}\isanewline
\isaindent{\ \ }\ \isakeywordONE{let}\ \isabound{vs}\ {\isacharequal}{\kern0pt}\ \isaconst{Field}\ {\isacharparenleft}{\kern0pt}\isaconst{ufa{\isacharunderscore}{\kern0pt}{\isasymalpha}}\ \isavar{uf}{\isacharparenright}{\kern0pt}\isanewline
\isaindent{\ \ \ }\isakeywordONE{in}\ {\isasymlparr}\isaconst{verts}\ {\isacharequal}{\kern0pt}\ \isabound{vs}{\isacharcomma}{\kern0pt}\isanewline
\isaindent{\ \ \ \isakeywordONE{in}\ \ \ \ }\isaconst{arcs}\ {\isacharequal}{\kern0pt}\ {\isacharbraceleft}{\kern0pt}{\isacharparenleft}{\kern0pt}\isaconst{ufa{\isacharunderscore}{\kern0pt}parent{\isacharunderscore}{\kern0pt}of}\ \isavar{uf}\ \isabound{x}{\isacharcomma}{\kern0pt}\ \isabound{x}{\isacharparenright}{\kern0pt}\ {\isacharbar}{\kern0pt}\ \isabound{x}\ {\isasymin}\ \isabound{vs}\ {\isasymand}\ \isaconst{ufa{\isacharunderscore}{\kern0pt}parent{\isacharunderscore}{\kern0pt}of}\ \isavar{uf}\ \isabound{x}\ {\isasymnoteq}\ \isabound{x}{\isacharbraceright}{\kern0pt}{\isasymrparr}}}\\[0.75em]
\isa{\isaconst{ufe{\isacharunderscore}{\kern0pt}forest{\isacharunderscore}{\kern0pt}of}\ \isafree{ufe}\ {\isasymequiv}\ \isaconst{ufa{\isacharunderscore}{\kern0pt}forest{\isacharunderscore}{\kern0pt}of}\ {\isacharparenleft}{\kern0pt}\isaconst{uf{\isacharunderscore}{\kern0pt}ds}\ \isafree{ufe}{\isacharparenright}{\kern0pt}}
\end{flushleft}
Note that we choose (somewhat arbitrarily) to direct the edges away from the root
because it aligns more naturally with the notion of a directed rooted tree.
Additionally, this choice ensures compatibility with the \isa{directed{\isacharunderscore}{\kern0pt}forest} locale,
which we implemented on top of the graph library.
For brevity, we omit the details here and direct the reader to the formalisation,
but suffice it to say that typical properties of forests,
e.g.\ the absence of cycles,
are proved in this locale.
To collect facts that are specific to \acrshort{uf} forests,
we define a locale \isa{ufa{\isacharunderscore}{\kern0pt}forest} fixing a \acrshort{uf} data structure \isa{\isafree{uf}}.
In the context of this locale,
we show that \isa{\isaconst{ufa{\isacharunderscore}{\kern0pt}forest{\isacharunderscore}{\kern0pt}of}\ \isafree{uf}} fulfils the requirements of a \isa{directed{\isacharunderscore}{\kern0pt}forest},
meaning that the facts in the latter locale transfer over to the former.
Similarly, we introduce the locale \isa{ufe{\isacharunderscore}{\kern0pt}forest} fixing a \acrshort{ufe} data structure \isa{\isafree{ufe}},
where \isa{\isaconst{uf{\isacharunderscore}{\kern0pt}ds}\ \isafree{ufe}} is a \isa{ufa{\isacharunderscore}{\kern0pt}forest}.%
\end{isamarkuptext}\isamarkuptrue%
\isadelimdocument
\endisadelimdocument
\isatagdocument
\isamarkupsubsubsection{Associated unions%
}
\isamarkuptrue%
\endisatagdocument
{\isafolddocument}%
\isadelimdocument
\endisadelimdocument
\begin{isamarkuptext}%
As illustrated by \cref{fig:explain'_abs},
we annotate each edge of the \acrshort{ufe} forest with the union that caused its creation,
i.e.\, for an effective union \isa{{\isacharparenleft}{\kern0pt}\isafree{a}{\isacharcomma}{\kern0pt}\ \isafree{b}{\isacharparenright}{\kern0pt}},
we annotate the newly created edge \isa{\isafree{e}} between the \isa{\isaconst{ufe{\isacharunderscore}{\kern0pt}rep{\isacharunderscore}{\kern0pt}of}\ \isafree{ufe}\ \isafree{a}} and \isa{\isaconst{ufe{\isacharunderscore}{\kern0pt}rep{\isacharunderscore}{\kern0pt}of}\ \isafree{ufe}\ \isafree{b}}
with \isa{{\isacharparenleft}{\kern0pt}\isafree{a}{\isacharcomma}{\kern0pt}\ \isafree{b}{\isacharparenright}{\kern0pt}}.
We say that \isa{{\isacharparenleft}{\kern0pt}\isafree{a}{\isacharcomma}{\kern0pt}\ \isafree{b}{\isacharparenright}{\kern0pt}} is the \emph{associated union} of \isa{\isafree{e}}.
Since the underlying \acrshort{uf} data structure is expressed in terms of parent pointers,
we actually associate the union \isa{{\isacharparenleft}{\kern0pt}\isafree{a}{\isacharcomma}{\kern0pt}\ \isafree{b}{\isacharparenright}{\kern0pt}} with \isa{\isaconst{ufe{\isacharunderscore}{\kern0pt}rep{\isacharunderscore}{\kern0pt}of}\ \isafree{ufe}\ \isafree{a}}.
Furthermore, we use an index into \isa{\isaconst{unions}\ \isafree{ufe}} rather than storing the union \isa{{\isacharparenleft}{\kern0pt}\isafree{a}{\isacharcomma}{\kern0pt}\ \isafree{b}{\isacharparenright}{\kern0pt}} directly.
This concept is formalised in the constant \isa{\isaconst{au{\isacharunderscore}{\kern0pt}ds}\ {\isacharcolon}{\kern0pt}{\isacharcolon}{\kern0pt}\ \isatconst{ufe}\ {\isasymRightarrow}\ \isatconst{nat}\ {\isasymRightarrow}\ \isatconst{nat}\ \isatconst{option}}
whose specific implementation we skip over here;
instead, we only state its characteristic properties:
\begin{itemize}
  \item \isa{{\normalsize{}If\,}\ \isaconst{unions}\ \isavar{ufe}\ {\isacharequal}{\kern0pt}\ {\isacharbrackleft}{\kern0pt}{\isacharbrackright}{\kern0pt}\ {\normalsize \,then\,}\ \isaconst{au{\isacharunderscore}{\kern0pt}ds}\ \isavar{ufe}\ {\isacharequal}{\kern0pt}\ {\isacharparenleft}{\kern0pt}{\isasymlambda}\isabound{x}{\isachardot}{\kern0pt}\ \isaconst{None}{\isacharparenright}{\kern0pt}{\isachardot}{\kern0pt}}
  \item For an effective union \isa{{\isacharparenleft}{\kern0pt}\isafree{a}{\isacharcomma}{\kern0pt}\ \isafree{b}{\isacharparenright}{\kern0pt}}, i.e\ if we have \isa{\isaconst{eff{\isacharunderscore}{\kern0pt}union}\ {\isacharparenleft}{\kern0pt}\isaconst{uf{\isacharunderscore}{\kern0pt}ds}\ \isavar{ufe}{\isacharparenright}{\kern0pt}\ \isafree{a}\ \isafree{b}},
     it holds that \isa{\isaconst{au{\isacharunderscore}{\kern0pt}ds}\ {\isacharparenleft}{\kern0pt}\isaconst{ufe{\isacharunderscore}{\kern0pt}union}\ \isavar{ufe}\ \isafree{a}\ \isafree{b}{\isacharparenright}{\kern0pt}\ {\isacharequal}{\kern0pt}\ {\isacharparenleft}{\kern0pt}\isaconst{au{\isacharunderscore}{\kern0pt}ds}\ \isavar{ufe}{\isacharparenright}{\kern0pt}{\isacharparenleft}{\kern0pt}\isaconst{ufe{\isacharunderscore}{\kern0pt}rep{\isacharunderscore}{\kern0pt}of}\ \isavar{ufe}\ \isafree{a}\ {\isasymmapsto}\ {\isacharbar}{\kern0pt}\isaconst{unions}\ \isavar{ufe}{\isacharbar}{\kern0pt}{\isacharparenright}{\kern0pt}},
     where \isa{{\isacharparenleft}{\kern0pt}\isafree{f}{\isacharparenleft}{\kern0pt}\isafree{x}\ {\isasymmapsto}\ \isafree{y}{\isacharparenright}{\kern0pt}{\isacharparenright}{\kern0pt}\ \isafree{z}\ {\isacharequal}{\kern0pt}\ {\isacharparenleft}{\kern0pt}\isakeywordONE{if}\ \isafree{z}\ {\isacharequal}{\kern0pt}\ \isafree{x}\ \isakeywordONE{then}\ \isaconst{Some}\ \isafree{y}\ \isakeywordONE{else}\ \isafree{f}\ \isafree{z}{\isacharparenright}{\kern0pt}}.
\end{itemize}%
\end{isamarkuptext}\isamarkuptrue%
\isadelimdocument
\endisadelimdocument
\isatagdocument
\isamarkupsubsubsection{Determining the \acrshort{lca} in a \acrshort{ufe} forest%
}
\isamarkuptrue%
\endisatagdocument
{\isafolddocument}%
\isadelimdocument
\endisadelimdocument
\isadeliminvisible
\endisadeliminvisible
\isataginvisible
\isakeywordONE{lemma}\isamarkupfalse%
\ lca{\isacharunderscore}{\kern0pt}ufa{\isacharunderscore}{\kern0pt}lca{\isacharcolon}{\kern0pt}\ \isanewline
\ \ {\isacartoucheopen}{\isacharbraceleft}{\kern0pt}x{\isacharcomma}{\kern0pt}\ y{\isacharbraceright}{\kern0pt}\ {\isasymsubseteq}\ Field\ {\isacharparenleft}{\kern0pt}ufa{\isacharunderscore}{\kern0pt}{\isasymalpha}\ uf{\isacharparenright}{\kern0pt}\ {\isasymLongrightarrow}\ ufa{\isacharunderscore}{\kern0pt}rep{\isacharunderscore}{\kern0pt}of\ uf\ x\ {\isacharequal}{\kern0pt}\ ufa{\isacharunderscore}{\kern0pt}rep{\isacharunderscore}{\kern0pt}of\ uf\ y\isanewline
\ \ {\isasymLongrightarrow}\ pre{\isacharunderscore}{\kern0pt}digraph{\isachardot}{\kern0pt}lca\ {\isacharparenleft}{\kern0pt}ufa{\isacharunderscore}{\kern0pt}forest{\isacharunderscore}{\kern0pt}of\ uf{\isacharparenright}{\kern0pt}\ {\isacharparenleft}{\kern0pt}ufa{\isacharunderscore}{\kern0pt}lca\ uf\ x\ y{\isacharparenright}{\kern0pt}\ x\ y{\isacartoucheclose}\isanewline
\ \ \isakeywordONE{by}\isamarkupfalse%
\ {\isacharparenleft}{\kern0pt}use\ ufa{\isacharunderscore}{\kern0pt}forest{\isachardot}{\kern0pt}lca{\isacharunderscore}{\kern0pt}ufa{\isacharunderscore}{\kern0pt}lca\ \isakeywordTWO{in}\ {\isacartoucheopen}unfold\ verts{\isacharunderscore}{\kern0pt}ufa{\isacharunderscore}{\kern0pt}forest{\isacharunderscore}{\kern0pt}of{\isacharcomma}{\kern0pt}\ simp{\isacartoucheclose}{\isacharparenright}{\kern0pt}\isanewline
\isanewline
\isakeywordONE{lemma}\isamarkupfalse%
\ ufa{\isacharunderscore}{\kern0pt}lca{\isacharunderscore}{\kern0pt}ufa{\isacharunderscore}{\kern0pt}union{\isacharcolon}{\kern0pt}\isanewline
{\isacartoucheopen}eff{\isacharunderscore}{\kern0pt}union\ uf\ a\ b\ {\isasymLongrightarrow}\ {\isacharbraceleft}{\kern0pt}x{\isacharcomma}{\kern0pt}\ y{\isacharbraceright}{\kern0pt}\ {\isasymsubseteq}\ Field\ {\isacharparenleft}{\kern0pt}ufa{\isacharunderscore}{\kern0pt}{\isasymalpha}\ uf{\isacharparenright}{\kern0pt}\ {\isasymLongrightarrow}\isanewline
\ \ ufa{\isacharunderscore}{\kern0pt}rep{\isacharunderscore}{\kern0pt}of\ {\isacharparenleft}{\kern0pt}ufa{\isacharunderscore}{\kern0pt}union\ uf\ a\ b{\isacharparenright}{\kern0pt}\ x\ {\isacharequal}{\kern0pt}\ ufa{\isacharunderscore}{\kern0pt}rep{\isacharunderscore}{\kern0pt}of\ {\isacharparenleft}{\kern0pt}ufa{\isacharunderscore}{\kern0pt}union\ uf\ a\ b{\isacharparenright}{\kern0pt}\ y\ {\isasymLongrightarrow}\isanewline
{\isacharparenleft}{\kern0pt}ufa{\isacharunderscore}{\kern0pt}lca\ {\isacharparenleft}{\kern0pt}ufa{\isacharunderscore}{\kern0pt}union\ uf\ a\ b{\isacharparenright}{\kern0pt}\ x\ y\ {\isacharequal}{\kern0pt}\isanewline
\ \ {\isacharparenleft}{\kern0pt}if\ ufa{\isacharunderscore}{\kern0pt}rep{\isacharunderscore}{\kern0pt}of\ uf\ x\ {\isacharequal}{\kern0pt}\ ufa{\isacharunderscore}{\kern0pt}rep{\isacharunderscore}{\kern0pt}of\ uf\ y\ then\ ufa{\isacharunderscore}{\kern0pt}lca\ uf\ x\ y\isanewline
\ \ else\ ufa{\isacharunderscore}{\kern0pt}rep{\isacharunderscore}{\kern0pt}of\ uf\ b{\isacharparenright}{\kern0pt}{\isacharparenright}{\kern0pt}\isanewline
{\isacartoucheclose}\ \isakeywordONE{by}\isamarkupfalse%
\ {\isacharparenleft}{\kern0pt}use\ ufa{\isacharunderscore}{\kern0pt}lca{\isacharunderscore}{\kern0pt}ufa{\isacharunderscore}{\kern0pt}union\ \isakeywordTWO{in}\ simp{\isacharunderscore}{\kern0pt}all{\isacharparenright}{\kern0pt}%
\endisataginvisible
{\isafoldinvisible}%
\isadeliminvisible
\endisadeliminvisible
\begin{isamarkuptext}%
The first auxiliary functions lists the elements on the path from the representative to some element.
Similarly to \isa{\isaconst{ufa{\isacharunderscore}{\kern0pt}rep{\isacharunderscore}{\kern0pt}of}}, this function is only well-defined for elements \isa{\isafree{x}\ {\isasymin}\ \isaconst{Field}\ {\isacharparenleft}{\kern0pt}\isaconst{ufa{\isacharunderscore}{\kern0pt}{\isasymalpha}}\ \isafree{uf}{\isacharparenright}{\kern0pt}} of a given \acrshort{uf} data structure \isa{\isafree{uf}}.
Now, let \isa{\isafree{px}} be the path from the representative of \isa{\isafree{x}} to \isa{\isafree{x}}
and \isa{\isafree{py}} be the path from \isa{\isafree{y}}'s representative to \isa{\isafree{y}}.
Then, every element of a common prefix of \isa{\isafree{px}} and \isa{\isafree{py}} is a common ancestor of \isa{\isafree{x}} and \isa{\isafree{y}} and
the \acrshort{lca} is exactly the last element of the longest common prefix of \isa{\isafree{px}} and \isa{\isafree{py}}. 
\begin{flushleft}
{\parindent0pt\isa{\isaconst{awalk{\isacharunderscore}{\kern0pt}verts{\isacharunderscore}{\kern0pt}from{\isacharunderscore}{\kern0pt}rep}\ {\isacharcolon}{\kern0pt}{\isacharcolon}{\kern0pt}\ \isatconst{ufa}\ {\isasymRightarrow}\ \isatconst{nat}\ {\isasymRightarrow}\ \isatconst{nat}\ \isatconst{list}}\\[\funheadersep]\isa{\isaconst{awalk{\isacharunderscore}{\kern0pt}verts{\isacharunderscore}{\kern0pt}from{\isacharunderscore}{\kern0pt}rep}\ \isavar{uf}\ \isavar{x}\ {\isacharequal}{\kern0pt}\isanewline
\isaindent{\ \ }\ \isakeywordONE{let}\ \isabound{px}\ {\isacharequal}{\kern0pt}\ \isaconst{ufa{\isacharunderscore}{\kern0pt}parent{\isacharunderscore}{\kern0pt}of}\ \isavar{uf}\ \isavar{x}\isanewline
\isaindent{\ \ \ }\isakeywordONE{in}\ \isakeywordONE{if}\ \isabound{px}\ {\isacharequal}{\kern0pt}\ \isavar{x}\ \isakeywordONE{then}\ {\isacharbrackleft}{\kern0pt}\isavar{x}{\isacharbrackright}{\kern0pt}\ \isakeywordONE{else}\ \isaconst{awalk{\isacharunderscore}{\kern0pt}verts{\isacharunderscore}{\kern0pt}from{\isacharunderscore}{\kern0pt}rep}\ \isavar{uf}\ \isabound{px}\ {\isacharat}{\kern0pt}\ {\isacharbrackleft}{\kern0pt}\isavar{x}{\isacharbrackright}{\kern0pt}}} \\[0.75em]
{\parindent0pt\isa{\isaconst{ufa{\isacharunderscore}{\kern0pt}lca}\ {\isacharcolon}{\kern0pt}{\isacharcolon}{\kern0pt}\ \isatconst{ufa}\ {\isasymRightarrow}\ \isatconst{nat}\ {\isasymRightarrow}\ \isatconst{nat}\ {\isasymRightarrow}\ \isatconst{nat}}\\[\funheadersep]\isa{\isaconst{ufa{\isacharunderscore}{\kern0pt}lca}\ \isavar{uf}\ \isavar{x}\ \isavar{y}\ {\isacharequal}{\kern0pt}\isanewline
\isaindent{\ \ }\ \isakeywordONE{let}\ \isabound{px}\ {\isacharequal}{\kern0pt}\ \isaconst{awalk{\isacharunderscore}{\kern0pt}verts{\isacharunderscore}{\kern0pt}from{\isacharunderscore}{\kern0pt}rep}\ \isavar{uf}\ \isavar{x}{\isacharsemicolon}{\kern0pt}\ \isabound{py}\ {\isacharequal}{\kern0pt}\ \isaconst{awalk{\isacharunderscore}{\kern0pt}verts{\isacharunderscore}{\kern0pt}from{\isacharunderscore}{\kern0pt}rep}\ \isavar{uf}\ \isavar{y}\isanewline
\isaindent{\ \ \ }\isakeywordONE{in}\ \isaconst{last}\ {\isacharparenleft}{\kern0pt}\isaconst{longest{\isacharunderscore}{\kern0pt}common{\isacharunderscore}{\kern0pt}prefix}\ \isabound{px}\ \isabound{py}{\isacharparenright}{\kern0pt}}}
\end{flushleft}
Again, we abbreviate \isa{\isaconst{ufe{\isacharunderscore}{\kern0pt}lca}\ \isafree{ufe}\ {\isasymequiv}\ \isaconst{ufa{\isacharunderscore}{\kern0pt}lca}\ {\isacharparenleft}{\kern0pt}\isaconst{uf{\isacharunderscore}{\kern0pt}ds}\ \isafree{ufe}{\isacharparenright}{\kern0pt}}.
It holds that \isa{\isaconst{ufa{\isacharunderscore}{\kern0pt}lca}} is indeed an \acrshort{lca}
provided that the arguments share the same representative
and thus are in the same tree of the forest.
For brevity, we omit the definition of \isa{\isaconst{lca}} here and refer to the formalisation instead.
\begin{lemma}\label{thm:lca_ufa_lca}
If \isa{{\isacharbraceleft}{\kern0pt}\isavar{x}{\isacharcomma}{\kern0pt}\ \isavar{y}{\isacharbraceright}{\kern0pt}\ {\isasymsubseteq}\ \isaconst{Field}\ {\isacharparenleft}{\kern0pt}\isaconst{ufa{\isacharunderscore}{\kern0pt}{\isasymalpha}}\ \isavar{uf}{\isacharparenright}{\kern0pt}} and \isa{\isaconst{ufa{\isacharunderscore}{\kern0pt}rep{\isacharunderscore}{\kern0pt}of}\ \isavar{uf}\ \isavar{x}\ {\isacharequal}{\kern0pt}\ \isaconst{ufa{\isacharunderscore}{\kern0pt}rep{\isacharunderscore}{\kern0pt}of}\ \isavar{uf}\ \isavar{y}}, then \isa{\isaconst{lca}\ {\isacharparenleft}{\kern0pt}\isaconst{ufa{\isacharunderscore}{\kern0pt}forest{\isacharunderscore}{\kern0pt}of}\ \isavar{uf}{\isacharparenright}{\kern0pt}\ {\isacharparenleft}{\kern0pt}\isaconst{ufa{\isacharunderscore}{\kern0pt}lca}\ \isavar{uf}\ \isavar{x}\ \isavar{y}{\isacharparenright}{\kern0pt}\ \isavar{x}\ \isavar{y}}.
\end{lemma}
We later prove key properties of \opexplain{} using the induction principle from \cref{thm:ufe_induct},
making it essential to understand how the auxiliary functions behave under effective unions.
The lemma below shows that \isa{\isaconst{ufa{\isacharunderscore}{\kern0pt}lca}} is invariant under a union \isa{{\isacharparenleft}{\kern0pt}\isafree{a}{\isacharcomma}{\kern0pt}\ \isafree{b}{\isacharparenright}{\kern0pt}} 
if its arguments share the same representative beforehand.
Otherwise, the union introduces an edge from the representative of \isa{\isafree{a}} to that of \isa{\isafree{b}},
connecting the trees that \isa{\isafree{x}} and \isa{\isafree{y}} belong to at their respective roots.
Due to the orientation of this new edge,
we know that the \acrshort{lca} of \isa{\isafree{x}} and \isa{\isafree{y}} must be the representative of \isa{\isafree{b}} after performing the union.
\begin{lemma}\label{thm:ufa_lca_ufa_union}
Assume \isa{\isaconst{eff{\isacharunderscore}{\kern0pt}union}\ \isavar{uf}\ \isavar{a}\ \isavar{b}} and \isa{{\isacharbraceleft}{\kern0pt}\isavar{x}{\isacharcomma}{\kern0pt}\ \isavar{y}{\isacharbraceright}{\kern0pt}\ {\isasymsubseteq}\ \isaconst{Field}\ {\isacharparenleft}{\kern0pt}\isaconst{ufa{\isacharunderscore}{\kern0pt}{\isasymalpha}}\ \isavar{uf}{\isacharparenright}{\kern0pt}}.
If \isa{\isaconst{ufa{\isacharunderscore}{\kern0pt}rep{\isacharunderscore}{\kern0pt}of}\ {\isacharparenleft}{\kern0pt}\isaconst{ufa{\isacharunderscore}{\kern0pt}union}\ \isavar{uf}\ \isavar{a}\ \isavar{b}{\isacharparenright}{\kern0pt}\ \isavar{x}\ {\isacharequal}{\kern0pt}\ \isaconst{ufa{\isacharunderscore}{\kern0pt}rep{\isacharunderscore}{\kern0pt}of}\ {\isacharparenleft}{\kern0pt}\isaconst{ufa{\isacharunderscore}{\kern0pt}union}\ \isavar{uf}\ \isavar{a}\ \isavar{b}{\isacharparenright}{\kern0pt}\ \isavar{y}} then \isa{\isaconst{ufa{\isacharunderscore}{\kern0pt}lca}\ {\isacharparenleft}{\kern0pt}\isaconst{ufa{\isacharunderscore}{\kern0pt}union}\ \isavar{uf}\ \isavar{a}\ \isavar{b}{\isacharparenright}{\kern0pt}\ \isavar{x}\ \isavar{y}\ {\isacharequal}{\kern0pt}\ {\isacharparenleft}{\kern0pt}\isakeywordONE{if}\ \isaconst{ufa{\isacharunderscore}{\kern0pt}rep{\isacharunderscore}{\kern0pt}of}\ \isavar{uf}\ \isavar{x}\ {\isacharequal}{\kern0pt}\ \isaconst{ufa{\isacharunderscore}{\kern0pt}rep{\isacharunderscore}{\kern0pt}of}\ \isavar{uf}\ \isavar{y}\ \isakeywordONE{then}\ \isaconst{ufa{\isacharunderscore}{\kern0pt}lca}\ \isavar{uf}\ \isavar{x}\ \isavar{y}\ \isakeywordONE{else}\ \isaconst{ufa{\isacharunderscore}{\kern0pt}rep{\isacharunderscore}{\kern0pt}of}\ \isavar{uf}\ \isavar{b}{\isacharparenright}{\kern0pt}}.
\end{lemma}%
\end{isamarkuptext}\isamarkuptrue%
\isadelimdocument
\endisadelimdocument
\isatagdocument
\isamarkupsubsubsection{Finding the most recent union on a path%
}
\isamarkuptrue%
\endisatagdocument
{\isafolddocument}%
\isadelimdocument
\endisadelimdocument
\begin{isamarkuptext}%
For the second auxiliary function,
we walk the path from the second argument \isa{\isafree{x}} to the first argument \isa{\isafree{y}}
and return the most recent associated union, i.e.\ the maximum index with respect to \isa{\isaconst{unions}\ \isafree{ufe}} on that path.
In Isabelle, we define the following function.
\begin{flushleft}
{\parindent0pt\isa{\isaconst{find{\isacharunderscore}{\kern0pt}newest{\isacharunderscore}{\kern0pt}on{\isacharunderscore}{\kern0pt}path}\ {\isacharcolon}{\kern0pt}{\isacharcolon}{\kern0pt}\ \isatconst{ufe}\ {\isasymRightarrow}\ \isatconst{nat}\ {\isasymRightarrow}\ \isatconst{nat}\ {\isasymRightarrow}\ \isatconst{nat}\ \isatconst{option}}\\[\funheadersep]\isa{\isaconst{find{\isacharunderscore}{\kern0pt}newest{\isacharunderscore}{\kern0pt}on{\isacharunderscore}{\kern0pt}path}\ \isavar{ufe}\ \isavar{y}\ \isavar{x}\ {\isacharequal}{\kern0pt}\isanewline
\isaindent{\ \ }\ \isakeywordONE{if}\ \isavar{y}\ {\isacharequal}{\kern0pt}\ \isavar{x}\ \isakeywordONE{then}\ \isaconst{None}\isanewline
\isaindent{\ \ \ }\isakeywordONE{else}\ \isaconst{max}\ {\isacharparenleft}{\kern0pt}\isaconst{au{\isacharunderscore}{\kern0pt}ds}\ \isavar{ufe}\ \isavar{x}{\isacharparenright}{\kern0pt}\ {\isacharparenleft}{\kern0pt}\isaconst{find{\isacharunderscore}{\kern0pt}newest{\isacharunderscore}{\kern0pt}on{\isacharunderscore}{\kern0pt}path}\ \isavar{ufe}\ \isavar{y}\ {\isacharparenleft}{\kern0pt}\isaconst{ufe{\isacharunderscore}{\kern0pt}parent{\isacharunderscore}{\kern0pt}of}\ \isavar{ufe}\ \isavar{x}{\isacharparenright}{\kern0pt}{\isacharparenright}{\kern0pt}}}\\[1em]
\end{flushleft}
As explained earlier, we only use this function on an element in conjunction with its \acrshort{lca} relative to another element.
Thus, there is a path between the two arguments and the function is well-defined for such inputs.
The path, however, can be empty, in which we return \isa{\isaconst{None}}, making the function partial.

As before, we are interested in how the function behaves under effective unions.
Since unions only join trees at their roots, existing paths in the tree are unchanged by unions,
so, for elements in the same equivalence class, the function is invariant under unions.
If, on the other hand, two elements only become part of the same equivalence class as a result of a union \isa{{\isacharparenleft}{\kern0pt}\isafree{a}{\isacharcomma}{\kern0pt}\ \isafree{b}{\isacharparenright}{\kern0pt}},
then \isa{{\isacharparenleft}{\kern0pt}\isafree{a}{\isacharcomma}{\kern0pt}\ \isafree{b}{\isacharparenright}{\kern0pt}} must be on the path between those elements
and, as it is the most recent union, the function returns the index of that union.
\begin{lemma}\label{thm:find_newest_on_path_ufe_union_if_reachable}
Assume that \isa{\isaconst{eff{\isacharunderscore}{\kern0pt}union}\ {\isacharparenleft}{\kern0pt}\isaconst{uf{\isacharunderscore}{\kern0pt}ds}\ \isavar{ufe}{\isacharparenright}{\kern0pt}\ \isavar{a}\ \isavar{b}}
and \isa{\isavar{y}\ {\normalsize \,is\ reachable\ from\,}\ \isavar{x}\ {\normalsize \,in\,}\ \isaconst{ufe{\isacharunderscore}{\kern0pt}forest{\isacharunderscore}{\kern0pt}of}\ {\isacharparenleft}{\kern0pt}\isaconst{ufe{\isacharunderscore}{\kern0pt}union}\ \isavar{ufe}\ \isavar{a}\ \isavar{b}{\isacharparenright}{\kern0pt}},
then it holds that \isa{\isaconst{find{\isacharunderscore}{\kern0pt}newest{\isacharunderscore}{\kern0pt}on{\isacharunderscore}{\kern0pt}path}\ {\isacharparenleft}{\kern0pt}\isaconst{ufe{\isacharunderscore}{\kern0pt}union}\ \isavar{ufe}\ \isavar{a}\ \isavar{b}{\isacharparenright}{\kern0pt}\ \isavar{x}\ \isavar{y}\ {\isacharequal}{\kern0pt}\ {\isacharparenleft}{\kern0pt}\isakeywordONE{if}\ \isaconst{ufe{\isacharunderscore}{\kern0pt}rep{\isacharunderscore}{\kern0pt}of}\ \isavar{ufe}\ \isavar{x}\ {\isacharequal}{\kern0pt}\ \isaconst{ufe{\isacharunderscore}{\kern0pt}rep{\isacharunderscore}{\kern0pt}of}\ \isavar{ufe}\ \isavar{y}\ \isakeywordONE{then}\ \isaconst{find{\isacharunderscore}{\kern0pt}newest{\isacharunderscore}{\kern0pt}on{\isacharunderscore}{\kern0pt}path}\ \isavar{ufe}\ \isavar{x}\ \isavar{y}\ \isakeywordONE{else}\ \isaconst{Some}\ {\isacharbar}{\kern0pt}\isaconst{unions}\ \isavar{ufe}{\isacharbar}{\kern0pt}{\isacharparenright}{\kern0pt}}.
\end{lemma}%
\end{isamarkuptext}\isamarkuptrue%
\isadelimdocument
\endisadelimdocument
\isatagdocument
\isamarkupsubsubsection{Explain%
}
\isamarkuptrue%
\endisatagdocument
{\isafolddocument}%
\isadelimdocument
\endisadelimdocument
\begin{isamarkuptext}%
With the auxiliary functions in place, we are set to implement the efficient \opexplain{} operation as shown in \cref{fig:explain'}.

Given arguments \isa{\isafree{x}} and \isa{\isafree{y}}, we first check whether they are equal
and, if so, we justify their equality by reflexivity.

Otherwise, we determine the \acrshort{lca} of the two elements and the most recent associated union on both of the paths from the elements to the \acrshort{lca}.
Note that, if the \acrshort{lca} is equal to \isa{\isafree{x}} or to \isa{\isafree{y}},
the respective path to the \acrshort{lca} is empty;
nevertheless, it is impossible that both \isa{\isafree{x}} and \isa{\isafree{y}} are equal to the \acrshort{lca}
because we are in the case where \isa{\isafree{x}\ {\isasymnoteq}\ \isafree{y}}.
Consider, for the sake of an explanation, the case where the most recent union \isa{{\isacharparenleft}{\kern0pt}\isafree{ax}{\isacharcomma}{\kern0pt}\ \isafree{bx}{\isacharparenright}{\kern0pt}} is on the path to \isa{\isafree{x}}.
This means, as illustrated in \cref{fig:explain'_abs}, that \isa{\isafree{x}} and \isa{\isafree{ax}}
as well as \isa{\isafree{y}} and \isa{\isafree{bx}}
are in the same subtree, respectively.
Thus, we call \isa{\isaconst{explain{\isacharprime}{\kern0pt}}} recursively and, using transitivity,
combine the resulting proofs of \isa{\isafree{x}\ {\isacharequal}{\kern0pt}\ \isafree{ax}} and \isa{\isafree{bx}\ {\isacharequal}{\kern0pt}\ \isafree{y}} with the assumption that \isa{\isafree{ax}\ {\isacharequal}{\kern0pt}\ \isafree{bx}}.

The last case, where the most recent union is on the path from \isa{\isafree{y}} to the \acrshort{lca},
is symmetric, which, accordingly, requires us to apply the symmetry rule after using the assumption rule on the most recent union.

As we will show below, \isa{\isaconst{explain{\isacharprime}{\kern0pt}}} only terminates for specific inputs.
The domain on which the function is well-defined is again characterised by a domain predicate
\isa{\isaconst{explain{\isacharprime}{\kern0pt}{\isacharunderscore}{\kern0pt}dom}\ {\isacharcolon}{\kern0pt}{\isacharcolon}{\kern0pt}\ \isatconst{ufe}\ {\isasymRightarrow}\ \isatconst{nat}\ {\isasymtimes}\ \isatconst{nat}\ {\isasymRightarrow}\ \isatconst{bool}}.
\begin{figure*}[t]
\begin{flushleft}
{\parindent0pt\isa{\isaconst{explain{\isacharprime}{\kern0pt}}\ {\isacharcolon}{\kern0pt}{\isacharcolon}{\kern0pt}\ \isatconst{ufe}\ {\isasymRightarrow}\ \isatconst{nat}\ {\isasymRightarrow}\ \isatconst{nat}\ {\isasymRightarrow}\ \isatconst{nat}\ \isatconst{eq{\isacharunderscore}{\kern0pt}prf}}\\[\funheadersep]\isa{\isaconst{explain{\isacharprime}{\kern0pt}}\ \isavar{ufe}\ \isavar{x}\ \isavar{y}\ {\isacharequal}{\kern0pt}\isanewline
\isaindent{\ \ }\ \isakeywordONE{if}\ \isavar{x}\ {\isacharequal}{\kern0pt}\ \isavar{y}\ \isakeywordONE{then}\ \isaconst{ReflP}\ \isavar{x}\isanewline
\isaindent{\ \ \ }\isakeywordONE{else}\ \isakeywordONE{let}\ \isabound{lca}\ {\isacharequal}{\kern0pt}\ \isaconst{ufe{\isacharunderscore}{\kern0pt}lca}\ \isavar{ufe}\ \isavar{x}\ \isavar{y}{\isacharsemicolon}{\kern0pt}\isanewline
\isaindent{\ \ \ \isakeywordONE{else}\ \isakeywordONE{let}\ }\isabound{newest{\isacharunderscore}{\kern0pt}x}\ {\isacharequal}{\kern0pt}\ \isaconst{find{\isacharunderscore}{\kern0pt}newest{\isacharunderscore}{\kern0pt}on{\isacharunderscore}{\kern0pt}path}\ \isavar{ufe}\ \isabound{lca}\ \isavar{x}{\isacharsemicolon}{\kern0pt}\isanewline
\isaindent{\ \ \ \isakeywordONE{else}\ \isakeywordONE{let}\ }\isabound{newest{\isacharunderscore}{\kern0pt}y}\ {\isacharequal}{\kern0pt}\ \isaconst{find{\isacharunderscore}{\kern0pt}newest{\isacharunderscore}{\kern0pt}on{\isacharunderscore}{\kern0pt}path}\ \isavar{ufe}\ \isabound{lca}\ \isavar{y}\isanewline
\isaindent{\ \ \ \isakeywordONE{else}\ }\isakeywordONE{in}\ \isakeywordONE{if}\ \isabound{newest{\isacharunderscore}{\kern0pt}y}\ {\isasymle}\ \isabound{newest{\isacharunderscore}{\kern0pt}x}\isanewline
\isaindent{\ \ \ \isakeywordONE{else}\ \isakeywordONE{in}\ }\isakeywordONE{then}\ \isakeywordONE{let}\ {\isacharparenleft}{\kern0pt}\isabound{ax}{\isacharcomma}{\kern0pt}\ \isabound{bx}{\isacharparenright}{\kern0pt}\ {\isacharequal}{\kern0pt}\ \isaconst{unions}\ \isavar{ufe}\ {\isacharbang}{\kern0pt}\ \isaconst{the}\ \isabound{newest{\isacharunderscore}{\kern0pt}x}\isanewline
\isaindent{\ \ \ \isakeywordONE{else}\ \isakeywordONE{in}\ \isakeywordONE{then}\ }\isakeywordONE{in}\ \isaconst{explain{\isacharprime}{\kern0pt}}\ \isavar{ufe}\ \isavar{x}\ \isabound{ax}\ \ensuremath{\bigtriangledown}\ \isaconst{AssmP}\ {\isacharparenleft}{\kern0pt}\isaconst{the}\ \isabound{newest{\isacharunderscore}{\kern0pt}x}{\isacharparenright}{\kern0pt}\ \ensuremath{\bigtriangledown}\isanewline
\isaindent{\ \ \ \isakeywordONE{else}\ \isakeywordONE{in}\ \isakeywordONE{then}\ \isakeywordONE{in}\ }\isaconst{explain{\isacharprime}{\kern0pt}}\ \isavar{ufe}\ \isabound{bx}\ \isavar{y}\isanewline
\isaindent{\ \ \ \isakeywordONE{else}\ \isakeywordONE{in}\ }\isakeywordONE{else}\ \isakeywordONE{let}\ {\isacharparenleft}{\kern0pt}\isabound{ay}{\isacharcomma}{\kern0pt}\ \isabound{by}{\isacharparenright}{\kern0pt}\ {\isacharequal}{\kern0pt}\ \isaconst{unions}\ \isavar{ufe}\ {\isacharbang}{\kern0pt}\ \isaconst{the}\ \isabound{newest{\isacharunderscore}{\kern0pt}y}\isanewline
\isaindent{\ \ \ \isakeywordONE{else}\ \isakeywordONE{in}\ \isakeywordONE{else}\ }\isakeywordONE{in}\ \isaconst{explain{\isacharprime}{\kern0pt}}\ \isavar{ufe}\ \isavar{x}\ \isabound{by}\ \ensuremath{\bigtriangledown}\ \isaconst{SymP}\ {\isacharparenleft}{\kern0pt}\isaconst{AssmP}\ {\isacharparenleft}{\kern0pt}\isaconst{the}\ \isabound{newest{\isacharunderscore}{\kern0pt}y}{\isacharparenright}{\kern0pt}{\isacharparenright}{\kern0pt}\ \ensuremath{\bigtriangledown}\isanewline
\isaindent{\ \ \ \isakeywordONE{else}\ \isakeywordONE{in}\ \isakeywordONE{else}\ \isakeywordONE{in}\ }\isaconst{explain{\isacharprime}{\kern0pt}}\ \isavar{ufe}\ \isabound{ay}\ \isavar{y}}}
\end{flushleft}
  \caption{Efficient version of the \opexplain{} operation.\label{fig:explain'}}
\end{figure*}%
\end{isamarkuptext}\isamarkuptrue%
\isadelimdocument
\endisadelimdocument
\isatagdocument
\isamarkupsubsection{Correctness\label{sec:explain'_correct}%
}
\isamarkuptrue%
\endisatagdocument
{\isafolddocument}%
\isadelimdocument
\endisadelimdocument
\begin{isamarkuptext}%
Verifying the functional correctness of \isa{\isaconst{explain{\isacharprime}{\kern0pt}}} requires
proving termination as well as soundness and completeness.
We prove termination directly, while we obtain soundness and completeness
by showing extensional equality of \isa{\isaconst{explain{\isacharprime}{\kern0pt}}} and \isa{\isaconst{explain}}.
As \isa{\isaconst{explain{\isacharprime}{\kern0pt}}}, like \isa{\isaconst{explain}}, does not validate its input,
we assume for the remainder of this \lcnamecref{sec:explain'_correct} that
\begin{enumerate*}
  \item \isa{{\isacharbraceleft}{\kern0pt}\isafree{x}{\isacharcomma}{\kern0pt}\ \isafree{y}{\isacharbraceright}{\kern0pt}\ {\isasymsubseteq}\ \isaconst{Field}\ {\isacharparenleft}{\kern0pt}\isaconst{ufe{\isacharunderscore}{\kern0pt}{\isasymalpha}}\ \isafree{ufe}{\isacharparenright}{\kern0pt}} and
  \item \isa{\isaconst{ufe{\isacharunderscore}{\kern0pt}rep{\isacharunderscore}{\kern0pt}of}\ \isafree{ufe}\ \isafree{x}\ {\isacharequal}{\kern0pt}\ \isaconst{ufe{\isacharunderscore}{\kern0pt}rep{\isacharunderscore}{\kern0pt}of}\ \isafree{ufe}\ \isafree{y}}.
\end{enumerate*}

To establish termination of \isa{\isaconst{explain{\isacharprime}{\kern0pt}}},
we first prove that termination remains invariant under an effective union
using the invariance of \isa{\isaconst{find{\isacharunderscore}{\kern0pt}newest{\isacharunderscore}{\kern0pt}on{\isacharunderscore}{\kern0pt}path}} and \isa{\isaconst{ufe{\isacharunderscore}{\kern0pt}lca}} under an effective union (see \cref{thm:ufa_lca_ufa_union,thm:find_newest_on_path_ufe_union_if_reachable}).
From this, the termination of \isa{\isaconst{explain{\isacharprime}{\kern0pt}}} follows by induction on \isa{\isatconst{ufe}}.
\begin{lemma}\label{thm:explain'_dom_ufe_union}
Assume \isa{\isaconst{explain{\isacharprime}{\kern0pt}{\isacharunderscore}{\kern0pt}dom}\ \isafree{ufe}\ {\isacharparenleft}{\kern0pt}\isafree{x}{\isacharcomma}{\kern0pt}\ \isafree{y}{\isacharparenright}{\kern0pt}}
and \isa{\isaconst{eff{\isacharunderscore}{\kern0pt}union}\ {\isacharparenleft}{\kern0pt}\isaconst{uf{\isacharunderscore}{\kern0pt}ds}\ \isafree{ufe}{\isacharparenright}{\kern0pt}\ \isavar{a}\ \isavar{b}},
then it holds that
\isa{\isaconst{explain{\isacharprime}{\kern0pt}{\isacharunderscore}{\kern0pt}dom}\ {\isacharparenleft}{\kern0pt}\isaconst{ufe{\isacharunderscore}{\kern0pt}union}\ \isafree{ufe}\ \isavar{a}\ \isavar{b}{\isacharparenright}{\kern0pt}\ {\isacharparenleft}{\kern0pt}\isafree{x}{\isacharcomma}{\kern0pt}\ \isafree{y}{\isacharparenright}{\kern0pt}}.
\end{lemma}
\begin{theorem}[Termination]\label{thm:explain'_dom}
\isa{\isaconst{explain{\isacharprime}{\kern0pt}{\isacharunderscore}{\kern0pt}dom}\ \isafree{ufe}\ {\isacharparenleft}{\kern0pt}\isafree{x}{\isacharcomma}{\kern0pt}\ \isafree{y}{\isacharparenright}{\kern0pt}}
\end{theorem}
By \cref{thm:explain'_dom} and the invariance of the auxiliary functions under effective unions,
we deduce that \isa{\isaconst{explain{\isacharprime}{\kern0pt}}} is also invariant under effective unions.
\begin{lemma}\label{thm:explain'_ufe_union}
\isa{{\normalsize{}If\,}\ \isaconst{eff{\isacharunderscore}{\kern0pt}union}\ {\isacharparenleft}{\kern0pt}\isaconst{uf{\isacharunderscore}{\kern0pt}ds}\ \isafree{ufe}{\isacharparenright}{\kern0pt}\ \isavar{a}\ \isavar{b}\ {\normalsize \,then\,}\ \isaconst{explain{\isacharprime}{\kern0pt}}\ {\isacharparenleft}{\kern0pt}\isaconst{ufe{\isacharunderscore}{\kern0pt}union}\ \isafree{ufe}\ \isavar{a}\ \isavar{b}{\isacharparenright}{\kern0pt}\ \isafree{x}\ \isafree{y}\ {\isacharequal}{\kern0pt}\ \isaconst{explain{\isacharprime}{\kern0pt}}\ \isafree{ufe}\ \isafree{x}\ \isafree{y}{\isachardot}{\kern0pt}}
\end{lemma}
Given the definition of \isa{\isaconst{explain}},
we now understand the behaviour of both \isa{\isaconst{explain}} and \isa{\isaconst{explain{\isacharprime}{\kern0pt}}} under effective unions.
Thus we conclude, by induction on \isa{\isatconst{ufe}}, that \isa{\isaconst{explain}} is extensionally equal to \isa{\isaconst{explain{\isacharprime}{\kern0pt}}}.
\begin{theorem}[Correctness]\label{thm:explain_eq_explain'}
\isa{\isaconst{explain}\ \isafree{ufe}\ \isafree{x}\ \isafree{y}\ {\isacharequal}{\kern0pt}\ \isaconst{explain{\isacharprime}{\kern0pt}}\ \isafree{ufe}\ \isafree{x}\ \isafree{y}}
\end{theorem}%
\end{isamarkuptext}\isamarkuptrue%
\isadelimdocument
\endisadelimdocument
\isatagdocument
\isamarkupsection{Refinement to an Efficiently Executable Specification\label{sec:refinement}%
}
\isamarkuptrue%
\endisatagdocument
{\isafolddocument}%
\isadelimdocument
\endisadelimdocument
\begin{isamarkuptext}%
In the previous section, we described a refined recursion scheme for \opexplain{} that avoids iterating through all input equalities.
To turn this into an efficiently executable specification, we refine two aspects of the \acrshort{ufe} data structure.

First, we employ the union-by-size heuristic~\cite{uf_by_size},
i.e.\ we always attach the tree with fewer elements to the one with more elements during a \opunion{}. 
This ensures that all trees in the \acrshort{uf} data structure have height at most $\mathcal{O}(\log n)$
where $n$ is the number of elements of the data structure.
This yields $\mathcal{O}(\log n)$ running time for \opunion{} and \opfind{} as well as $\mathcal{O}(k \log n)$ for \opexplain{}.

Then, we take this functional \acrshort{ufe} data structure
and refine it to an imperative specification, thereby giving a concrete implementation.
In doing that, we are careful to refine lists by arrays,
guaranteeing constant time access to e.g.\ the parent of an element in the \acrshort{uf} data structure.
Additionally, we maintain a copy of the \acrshort{uf} data structure with path compression
as described in \cref{sec:ufe_efficient},
improving the performance of \opunion{} and \opfind{} to almost constant running time.%
\end{isamarkuptext}\isamarkuptrue%
\isadelimdocument
\endisadelimdocument
\isatagdocument
\isamarkupsubsection{Union-by-size Heuristic%
}
\isamarkuptrue%
\endisatagdocument
{\isafolddocument}%
\isadelimdocument
\endisadelimdocument
\begin{isamarkuptext}%
As mentioned in \cref{sec:uf_hol}, our formalisation of the \acrshort{uf} data structure extends a formalisation by \citeauthor{uf_isabelle}~\cite{uf_isabelle,uf_isabelle_afp}.
The latter formalisation already introduces the union-by-size heuristic,
but it does so during the refinement to Imperative HOL.
We raise the union-by-size heuristic to the purely functional level of HOL,
which lets us exploit Isabelle's lifting and transfer infrastructure~\cite{lifting_transfer}.
In addition, we introduce another optimisation: we represent the \acrshort{uf} data structure as a single list of integers,
eliminating the data structure recording the size information.

As a prerequisite for the union-by-size heuristic,
we define a function that determines the equivalence class of an element \isa{\isafree{x}} in the data structure \isa{\isafree{uf}}.
More specifically, we use the relational image operator \isa{{\isacharparenleft}{\kern0pt}{\isacharbackquote}{\kern0pt}{\isacharbackquote}{\kern0pt}\:{\isacharparenright}{\kern0pt}} on the equivalence relation \isa{\isaconst{ufa{\isacharunderscore}{\kern0pt}{\isasymalpha}}\ \isafree{uf}}
to obtain all the elements that are equivalent to \isa{\isafree{x}}.
The associated size of an element is then the cardinality of its equivalence class.
\begin{flushleft}
\begin{minipage}{0.48\linewidth}
{\parindent0pt\isa{\isaconst{ufa{\isacharunderscore}{\kern0pt}eq{\isacharunderscore}{\kern0pt}class}\ {\isacharcolon}{\kern0pt}{\isacharcolon}{\kern0pt}\ \isatconst{ufa}\ {\isasymRightarrow}\ \isatconst{nat}\ {\isasymRightarrow}\ \isatconst{nat}\ \isatconst{set}}\\[\funheadersep]\isa{\isaconst{ufa{\isacharunderscore}{\kern0pt}eq{\isacharunderscore}{\kern0pt}class}\ \isavar{uf}\ \isafree{x}\ {\isacharequal}{\kern0pt}\ \isaconst{ufa{\isacharunderscore}{\kern0pt}{\isasymalpha}}\ \isavar{uf}\ {\isacharbackquote}{\kern0pt}{\isacharbackquote}{\kern0pt}\:\ {\isacharbraceleft}{\kern0pt}\isafree{x}{\isacharbraceright}{\kern0pt}}}
\end{minipage}
\hfill
\begin{minipage}{0.48\linewidth}
{\parindent0pt\isa{\isaconst{ufa{\isacharunderscore}{\kern0pt}size}\ {\isacharcolon}{\kern0pt}{\isacharcolon}{\kern0pt}\ \isatconst{ufa}\ {\isasymRightarrow}\ \isatconst{nat}\ {\isasymRightarrow}\ \isatconst{nat}}\\[\funheadersep]\isa{\isaconst{ufa{\isacharunderscore}{\kern0pt}size}\ \isavar{uf}\ \isafree{x}\ {\isacharequal}{\kern0pt}\ {\isacharbar}{\kern0pt}\isaconst{ufa{\isacharunderscore}{\kern0pt}eq{\isacharunderscore}{\kern0pt}class}\ \isavar{uf}\ \isafree{x}{\isacharbar}{\kern0pt}}}
\end{minipage}
\end{flushleft}
With this, we perform the \opunion{} operation such that the element with the smaller size is always passed as the first argument. 
The underlying implementation of the data structure always updates the parent pointer of the representative of the first argument to the representative of the second argument,
thus yielding a \opunion{} operation that attaches smaller trees in the \acrshort{uf} forest to larger trees.
\begin{flushleft}
{\parindent0pt\isa{\isaconst{ufa{\isacharunderscore}{\kern0pt}union{\isacharunderscore}{\kern0pt}size}\ {\isacharcolon}{\kern0pt}{\isacharcolon}{\kern0pt}\ \isatconst{ufa}\ {\isasymRightarrow}\ \isatconst{nat}\ {\isasymRightarrow}\ \isatconst{nat}\ {\isasymRightarrow}\ \isatconst{ufa}}\\[\funheadersep]\isa{\isaconst{ufa{\isacharunderscore}{\kern0pt}union{\isacharunderscore}{\kern0pt}size}\ \isavar{ufa}\ \isavar{x}\ \isavar{y}\ {\isacharequal}{\kern0pt}\isanewline
\isaindent{\ \ }\ \isakeywordONE{let}\ \isabound{rep{\isacharunderscore}{\kern0pt}x}\ {\isacharequal}{\kern0pt}\ \isaconst{ufa{\isacharunderscore}{\kern0pt}rep{\isacharunderscore}{\kern0pt}of}\ \isavar{ufa}\ \isavar{x}{\isacharsemicolon}{\kern0pt}\ \isabound{rep{\isacharunderscore}{\kern0pt}y}\ {\isacharequal}{\kern0pt}\ \isaconst{ufa{\isacharunderscore}{\kern0pt}rep{\isacharunderscore}{\kern0pt}of}\ \isavar{ufa}\ \isavar{y}\isanewline
\isaindent{\ \ \ }\isakeywordONE{in}\ \isakeywordONE{if}\ \isaconst{ufa{\isacharunderscore}{\kern0pt}size}\ \isavar{ufa}\ \isabound{rep{\isacharunderscore}{\kern0pt}x}\ {\isacharless}{\kern0pt}\ \isaconst{ufa{\isacharunderscore}{\kern0pt}size}\ \isavar{ufa}\ \isabound{rep{\isacharunderscore}{\kern0pt}y}\ \isakeywordONE{then}\ \isaconst{ufa{\isacharunderscore}{\kern0pt}union}\ \isavar{ufa}\ \isavar{x}\ \isavar{y}\isanewline
\isaindent{\ \ \ \isakeywordONE{in}\ }\isakeywordONE{else}\ \isaconst{ufa{\isacharunderscore}{\kern0pt}union}\ \isavar{ufa}\ \isavar{y}\ \isavar{x}}}
\end{flushleft}
Looking closely at the definition,
we see that \isa{\isaconst{ufa{\isacharunderscore}{\kern0pt}size}} is only ever used on the representative of an element.
Moreover, in the representation of \isa{\isatconst{ufa}} as a list of natural numbers,
the representatives are exactly those where the parent pointer is self-referential.
Ultimately, we integrate both insights and encode the \acrshort{uf} data structure
as an \acrshort{adt} \isa{\isatconst{ufsi}}, which is implemented by a list of integers:
we use a negative number to indicate that a parent pointer is self-referential,
using the absolute value of the number as the size at the same time.
The other parent pointers are encoded as non-negative numbers as before.%
\end{isamarkuptext}\isamarkuptrue%
\isadelimdocument
\endisadelimdocument
\isatagdocument
\isamarkupsubsection{From Functional to Imperative Specification\label{sec:imperative_hol}%
}
\isamarkuptrue%
\endisatagdocument
{\isafolddocument}%
\isadelimdocument
\endisadelimdocument
\begin{isamarkuptext}%
To obtain an imperative specification,
we formulate a refined version of the \opexplain{} operation in the heap monad provided by the Imperative HOL~\cite{imperative_hol} framework.
This framework comes with an extension to Isabelle's code generator allowing us to generate imperative code in several target languages including \acrlong{sml}. 
Since Imperative HOL only comes with limited capabilities to analyse programs in its heap monad, 
we bring in \citeauthor{uf_isabelle}'s~\cite{uf_isabelle} separation logic framework for Imperative HOL.
The framework lets us reason about the state of the heap using heap assertions,
which describe data stored on the heap and their properties.
All our data structures are ultimately represented as arrays on the heap,
so we ensure with heap assertions that the content of the arrays represents our data structures throughout the operations we perform on them. 

With the automation provided by \citeauthor{uf_isabelle}'s framework,
it is straightforward to implement the operations and prove their correctness.
The process is similar to the refinement of the \acrshort{uf} data structure~\cite{uf_isabelle}.
Thus, we forgo a discussion of how individual functions are refined and refer to the formalisation instead.

The only remaining noteworthy detail is the representation of the \acrshort{ufe} data structure in Imperative HOL.
Our implementation consists of a \acrshort{uf} data structure,
a partial function recording the associated union of each parent pointer,
and the chronological list of unions.
The \acrshort{uf} data structure is represented as an array of integers.
For the associated unions, we use an array of options to represent the partial function.
This works as the domain is actually fixed,
i.e.\ the domain of the partial function is exactly the elements of the \acrshort{uf} data structure,
which, in our case, are the natural numbers up to some fixed \isa{\isafree{n}}.
Lastly, we represent the list of unions as a dynamic array using the type \isa{array{\isacharunderscore}{\kern0pt}list}.
The type wraps an array together with a natural number indicating how many cells of the array,
counting from the first position,
are occupied.
We can then grow the array dynamically by pushing elements to the end,
doubling its size each time it becomes fully occupied.
Hence, we achieve amortised constant running time for adding new unions and constant time random access,
which are the operations required by the \opexplain{} operation. 
We assemble these components into a record type \isa{\isatconst{ufe{\isacharunderscore}{\kern0pt}imp}}.
Finally, we extend \isa{\isatconst{ufe{\isacharunderscore}{\kern0pt}imp}} with a \acrshort{uf} data structure with path compression,
thus obtaining the record type \isa{\isatconst{ufe{\isacharunderscore}{\kern0pt}c{\isacharunderscore}{\kern0pt}imp}}.

We define a heap assertion \isa{\isaconst{is{\isacharunderscore}{\kern0pt}ufe}\ {\isacharcolon}{\kern0pt}{\isacharcolon}{\kern0pt}\ \isatconst{ufe}\ {\isasymtimes}\ \isatconst{nat}\ {\isasymRightarrow}\ \isatconst{ufe{\isacharunderscore}{\kern0pt}imp}\ {\isasymRightarrow}\ \isatconst{assn}}
that relates instances of the \acrshort{adt} \isa{\isatconst{ufe}} with instances of \isa{\isatconst{ufe{\isacharunderscore}{\kern0pt}imp}}.
The assertion just relates the components of \isa{\isatconst{ufe{\isacharunderscore}{\kern0pt}imp}} with the corresponding functions on \isa{\isatconst{ufe}},
so we omit it for brevity.
The only aspect requiring further explanation is the natural number \isa{\isafree{n}} in the first argument.
Its purpose is to ensure that the elements of the initial \acrshort{uf} data structure
and the domain of the associated unions are both the numbers up to \isa{\isafree{n}}.
To obtain the assertion \isa{\isaconst{is{\isacharunderscore}{\kern0pt}ufe{\isacharunderscore}{\kern0pt}c}\ {\isacharcolon}{\kern0pt}{\isacharcolon}{\kern0pt}\ \isatconst{ufe}\ {\isasymtimes}\ \isatconst{nat}\ {\isasymRightarrow}\ \isatconst{ufe{\isacharunderscore}{\kern0pt}c{\isacharunderscore}{\kern0pt}imp}\ {\isasymRightarrow}\ \isatconst{assn}},
we additionally require that the representatives in the \acrshort{uf} data structure with path compression
corresponds to the representatives in the \acrshort{ufe} data structure.

Again, refining the operations on \isa{\isatconst{ufe{\isacharunderscore}{\kern0pt}c{\isacharunderscore}{\kern0pt}imp}} is routine;
so, we only show the final correctness theorem for \isa{\isaconst{explain{\isacharunderscore}{\kern0pt}partial{\isacharunderscore}{\kern0pt}imp}},
an imperative version of \isa{\isaconst{explain{\isacharprime}{\kern0pt}}} that ensures soundness
following the approach of \isa{\isaconst{explain{\isacharunderscore}{\kern0pt}partial}} in \cref{sec:ufe_simple}.
\begin{theorem}
We prove the following Hoare triple, which entails total correctness in the Separation Logic Framework~\cite{uf_isabelle_afp}:
\isa{{\isacharless}{\kern0pt}\isaconst{is{\isacharunderscore}{\kern0pt}ufe{\isacharunderscore}{\kern0pt}c}\ {\isacharparenleft}{\kern0pt}\isavar{ufe}{\isacharcomma}{\kern0pt}\ \isavar{n}{\isacharparenright}{\kern0pt}\ \isavar{ufe{\isacharunderscore}{\kern0pt}c{\isacharunderscore}{\kern0pt}imp}{\isachargreater}{\kern0pt}\ \isaconst{explain{\isacharunderscore}{\kern0pt}partial{\isacharunderscore}{\kern0pt}imp}\ \isavar{ufe{\isacharunderscore}{\kern0pt}c{\isacharunderscore}{\kern0pt}imp}\ \isavar{x}\ \isavar{y}\ {\isacharless}{\kern0pt}{\isasymlambda}\isabound{r}{\isachardot}{\kern0pt}\ \isaconst{is{\isacharunderscore}{\kern0pt}ufe{\isacharunderscore}{\kern0pt}c}\ {\isacharparenleft}{\kern0pt}\isavar{ufe}{\isacharcomma}{\kern0pt}\ \isavar{n}{\isacharparenright}{\kern0pt}\ \isavar{ufe{\isacharunderscore}{\kern0pt}c{\isacharunderscore}{\kern0pt}imp}\ {\isacharasterisk}{\kern0pt}\ {\isasymup}\ {\isacharparenleft}{\kern0pt}\isabound{r}\ {\isacharequal}{\kern0pt}\ \isaconst{explain{\isacharunderscore}{\kern0pt}partial}\ \isavar{ufe}\ \isavar{x}\ \isavar{y}{\isacharparenright}{\kern0pt}{\isachargreater}{\kern0pt}\isactrlsub t}
\end{theorem}%
\end{isamarkuptext}\isamarkuptrue%
\isadelimdocument
\endisadelimdocument
\isatagdocument
\isamarkupsection{Benchmarking the Exported Code%
}
\isamarkuptrue%
\endisatagdocument
{\isafolddocument}%
\isadelimdocument
\endisadelimdocument
\begin{isamarkuptext}%
In the previous section, we obtained an executable imperative specification of the \acrshort{ufe} data structure,
from which we can export code to functional target languages while
utilising their respective support for imperative programming like destructive array updates in \acrfull{sml}.
This raises the question whether exporting imperative code to a functional language is a good fit, performance wise. 
In addition, \acrshort{smt} solvers are usually implemented in imperative language such as \Cpp{}.
Therefore, we compare the exported \acrshort{sml} code against
a hand-written \Cpp{} implementation of the executable specification.

We analyse the performance on two test cases:
\begin{enumerate*}
  \item in the former case, the the number of proof steps for an \opexplain{} operation is linear in the number of elements but the
    depth of the \acrshort{uf} forest is constant,
  \item while in the latter, the depth of the \acrshort{uf} forest as well as the number of proof steps
    is logarithmic in the number of elements.
\end{enumerate*}
For both cases, we choose a natural number $n$, initialise the \acrshort{ufe} data structure with
$2^n$ elements, perform \opunion{} operations that results in the desired \acrshort{uf} forest,
and finally perform a number (i.e.\ 1000 and 100000, respectively) of \opexplain{} operations with the arguments drawn from
the uniform distribution over $0,\ldots,2^n - 1$.
We identify the test cases by functions \isa{\isaconst{wide}} and \isa{\isaconst{balanced}},
which both have type \isa{\isatconst{nat}\ {\isasymRightarrow}\ {\isacharparenleft}{\kern0pt}\isatconst{nat}\ {\isasymtimes}\ \isatconst{nat}{\isacharparenright}{\kern0pt}\ \isatconst{list}}.
\cref{fig:test_cases} illustrates the resulting \acrshort{uf} forests or, more specifically, trees.

\begin{figure}
\centering
\begin{subfigure}[b]{0.39\textwidth}
\centering
\begin{tikzpicture}[
  every node/.append style={draw, ellipse, minimum width=2em},
  sibling distance=1.2cm,
  >=stealth, edge from parent/.append style={draw, <-}]
  \node {$0$}
    child {node (1) {$1$}}
    child {node[draw=none] {$\ldots$}}
    child {node {$2^n - 1$}};
\end{tikzpicture}
\subcaption{\acrshort{uf} tree for \isa{\isaconst{wide}\ \isafree{n}}.}
\end{subfigure}
\hfill
\begin{subfigure}[b]{0.58\textwidth}
\centering
\begin{tikzpicture}[
  every node/.append style={draw, circle},
  sibling distance=0.9cm, level distance = 0.725cm,
  >=stealth, edge from parent/.append style={draw, <-}]
  \node (2) {$0$}
    child {node (21) {$1$}}
    child {node {$2$}
      child {node {$3$}}
      child[missing] {}
    }
    child {node {$4$}
      child {node {$5$}}
      child {node {$6$}
        child {node {$7$}}
        child[missing] {}
      }
    };
  \node[left=2cm of 2] (1) {$0$}
    child {node (11) {$1$}}
    child {node {$2$}
      child {node {$3$}}
      child[missing] {}
    };
  \node[left=1cm of 1] (0) {$0$}
    child {node (01) {$1$}}
    child[missing] {};
\end{tikzpicture}
\subcaption{\acrshort{uf} trees for \isa{\isaconst{balanced}\ \isafree{n}} where \isa{\isafree{n}\ {\isasymin}\ {\isacharbraceleft}{\kern0pt}{\isadigit{1}}{\isacharcomma}{\kern0pt}\ {\isadigit{2}}{\isacharcomma}{\kern0pt}\ {\isadigit{3}}{\isacharbraceright}{\kern0pt}}.}
\end{subfigure}
\caption{The \acrshort{uf} trees resulting from performing the \opunion{} operations with the arguments given by \isa{\isaconst{wide}} and \isa{\isaconst{balanced}}.\label{fig:test_cases}}
\end{figure}

To perform our measurements, we compiled the exported Standard ML code with MLton\footnote{\url{http://mlton.org}} version \texttt{20210117},
and the \Cpp{} code with \Gpp{}\footnote{\url{https://gcc.gnu.org}} version \texttt{13.3.0}.
The results are shown in \cref{tab:benchmark}.
The code export of Isabelle uses arbitrary sized integers to ensure soundness with respect to the executable specification
while the \Cpp{} uses machine integers,
so we also include a version of the exported code, annotated by (Int), that uses machine integers.

The observed running time overhead of using arbitrary sized integers is roughly a factor of 2--2.5,
matching that observed by \citeauthor{refine_monadic}~\cite{refine_monadic}.
The difference between \acrshort{sml} with machine integers and \Cpp{} is
roughly a factor of 2 for the \opunion{} operations and
a factor of 1.5 for the \opexplain{} operations throughout both test cases.
The second test case exhibits some outliers:
notably, between $n = 23$ to $n = 24$ for \acrshort{sml} and
between $n = 24$ and $n = 25$ for \acrshort{sml} (Int).
This variance is due to garbage collection becoming a significant factor at large heap sizes,
e.g. for $n = 25$ the heap grows to above 5 gigabytes.
\begin{table}[t]
\centering
\begin{subtable}{\textwidth}
\centering
\begin{tabular}{l@ {\hskip 6pt} l@ {\hskip 6pt}l@ {\hskip 6pt}l@ {\hskip 6pt}l@ {\hskip 6pt}l}
\toprule
Impl. & 18 & 19 & 20 & 21 & 22\\
\midrule
SML & 0.025/18.428 & 0.075/36.129 & 0.072/70.696 & 0.157/140.209 & 0.393/280.684 \\
SML (Int) & 0.011/8.341 & 0.011/16.249 & 0.024/29.695 & 0.051/62.442 & 0.092/131.532 \\
\Cpp{} & 0.004/3.672 & 0.007/7.354 & 0.015/15.113 & 0.031/31.120 & 0.062/71.066 \\
\bottomrule
\end{tabular}
\subcaption{Running times for \isa{\isaconst{wide}\ \isafree{n}}.}
\end{subtable}

\begin{subtable}{\textwidth}
\centering
\begin{tabular}{l@ {\hskip 6pt}l@ {\hskip 6pt}l@ {\hskip 6pt}l@ {\hskip 6pt}l@ {\hskip 6pt}l}
\toprule
Impl. & 22 & 23 & 24 & 25 & 26\\
\midrule
SML & 0.722/1.879 & 0.899/2.583 & 1.552/2.082 & 4.770/2.396 & 14.610/3.014 \\
SML (Int) & 0.174/0.750 & 0.227/0.781 & 0.695/0.900 & 2.474/0.920 & 2.785/1.027 \\
\Cpp{} & 0.087/0.369 & 0.199/0.460 & 0.350/0.550 & 0.752/0.620 & 1.451/0.728 \\
\bottomrule
\end{tabular}
\subcaption{Running times for \isa{\isaconst{balanced}\ \isafree{n}}.}
\end{subtable}
\caption{
  Wall-clock running times in seconds as measured on an Intel Core i7 4790k.
  For each $n$, we recorded the running time for performing the \opunion{} operations and the \opexplain{} operations (separated by a slash). \label{tab:benchmark}}
\end{table}

Overall, we found that employing a functional language results in a modest performance overhead when working with machine integers.
We note that to soundly export such code,
it would be necessary to change the element type of the \acrshort{ufe} data structure from natural numbers to fixed-width words.
This is plausible future work, as the number of elements is fixed for any instance of the data structure,
and the only necessary operations on the elements are comparisons and indexing into arrays --- operations that fixed-width words also support.%
\end{isamarkuptext}\isamarkuptrue%
\isadelimdocument
\endisadelimdocument
\isatagdocument
\isamarkupsection{Conclusion and Future Work%
}
\isamarkuptrue%
\endisatagdocument
{\isafolddocument}%
\isadelimdocument
\endisadelimdocument
\begin{isamarkuptext}%
We developed a formalisation of the \acrshort{uf} data structure with an \opexplain{} operation 
based on a paper by \citeauthor{congcl_proofs}~\cite{congcl_proofs}.
The formalisation includes a more naive version of the \opexplain{} operation than the one presented in the paper.
We proved their equivalence as well as their soundness, completeness, and termination.
Finally, we refined the functional representation of the data structure to an imperative one, allowing us to export efficient code.

In future work, we plan to verify the other variant of the \acrshort{ufe} data structure as presented by \citeauthor{congcl_proofs}.
This variant also forms the basis of their congruence closure algorithm, which is the logical next step.
Ultimately, we want to work towards a verified, proof-producing version of the Nelson-Oppen algorithm~\cite{nelson_oppen} for the combination of theories.

\subsubsection*{Acknowledgements}
We thank Tobias Nipkow for reviewing a draft version of this paper and the anonymous referees for their thoughtful feedback.

{\fontsize{9}{15}\selectfont
\subsubsection*{\fontsize{9}{15}\selectfont Disclosure of Interests}
The authors have no competing interests to declare that are relevant to the content of this article.
}%
\end{isamarkuptext}\isamarkuptrue%
\begin{isamarkuptext}%
\bibliographystyle{splncs04nat}
\bibliography{root}%
\end{isamarkuptext}\isamarkuptrue%
\isadelimtheory
\endisadelimtheory
\isatagtheory
\isakeywordTWO{end}\isamarkupfalse%
\endisatagtheory
{\isafoldtheory}%
\isadelimtheory
\endisadelimtheory
\end{isabellebody}%

\end{document}